\documentclass[aos]{imsart}

\RequirePackage{amsthm,amsmath,amsfonts,amssymb}

\RequirePackage[numbers,sort&compress]{natbib}
\RequirePackage[colorlinks,citecolor=blue,urlcolor=blue]{hyperref}
\RequirePackage{graphicx}

\usepackage{lmodern}
\usepackage{comment}

\let\large=\normalsize

\let\Huge=\large
\let\huge=\large
\let\LARGE=\large
\let\Large=\large
\newtheorem{theorem}{Theorem}
\newtheorem{lemma}{Lemma}

\newtheorem{corollary}{Corollary}

\let\tilde=\widetilde
\let\hat=\widehat

\newcommand{\eps}{\varepsilon}

\DeclareMathOperator*{\diag}{diag}

\DeclareMathOperator*{\argmax}{argmax}

\def\var{\mbox{var}}

\def\tr{\mbox{tr}}

\def\sumtT{\sum_{t=1}^T}
\def\sumiN{\sum_{i=1}^N}
\def\qed{\hfill $\square$}

\newcommand{\N}{\mathcal{N}}

\newcommand{\bbig}{\Big}
\newcommand{\convd}{\stackrel{d}{\longrightarrow}}

\newcommand{\DD}{D}

\newcommand{\HK}{Hahn and Kuersteiner }

\newcommand{\R}{R}

\renewcommand{\tr}{\mathrm{tr}}

\newcommand{\rootNT}{\frac 1 {\sqrt{NT} }  }
\newcommand{\tF}{\tilde F}

\newcommand{\FplustF}{F+\rootNT \tF}
\newcommand{\talpha}{\tilde \alpha}
\newcommand{\op}{o_p(1)}

\newcommand{\ttheta}{\tilde \theta}

\allowdisplaybreaks[2]

\newcommand{\rootT}{\frac 1 {\sqrt{T}}}
\newcommand{\rootN}{\frac 1 {\sqrt{N}}}
\renewcommand{\R}{\mathbb R}
\renewcommand{\H}{\mathbb H}
\newcommand{\tf}{\tilde f}
\newcommand{\ta}{\tilde \alpha}

\newcommand{\E}{\mathbb E}
\newcommand{\tpsi}{\tilde \psi}
\newcommand{\FFD}{(FF'+D)^{-1}}
\newcommand{\DMDFD}{D^{-1/2} M_{D^{-1/2}F} D^{-1/2}}

\begin{document}

\begin{frontmatter}

\title{ Efficiency of QMLE for dynamic panel data models with interactive effects}
\runtitle{Dynamic panel data models}

\begin{aug}
\author[A]{\fnms{Jushan}~\snm{Bai}\ead[label=e1]{Jushan.bai@columbia.edu}},

\address[A]{Department of Economics,
Columbia University\printead[presep={,\ }]{e1}}

\end{aug}

\begin{abstract}

This paper studies the problem of efficient estimation of panel data models in
the presence of an increasing number of incidental parameters.
We formulate the dynamic panel as a  simultaneous equations system, and derive the efficiency bound
under the normality assumption. We  then  show that the Gaussian quasi-maximum likelihood estimator
(QMLE) applied to the system  achieves the normality efficiency bound  without the normality assumption.
Comparison of QMLE with the fixed effects approach is made.

\end{abstract}



\end{frontmatter}

\section{Introduction}

Consider the dynamic panel data model with interactive effects
\begin{equation} \label{eq:dyn-interative} y_{it}=\alpha \, y_{it-1}+\delta_t + \lambda_i'f_t + \eps_{it} \end{equation}
\[ i=1,2,...,N;  t=1,...,T\]
where $y_{it}$ is the outcome variable, $\lambda_i$
and $f_t$ are each $r\times 1$ and both are unobservable, $\delta_t$ is the
time effect, and $\eps_{it}$ is the error term.   Only $y_{it}$ are observable.
The above model is increasingly used for empirical studies in social sciences. The purpose of this paper is about efficient estimation of the model by deriving the efficiency bound.
We show that quasi-maximum likelihood estimation (QMLE) achieves the  efficiency bound.  But first, we explain the meaning of QMLE  for this model and its motivations.

The index $i$ is referred to as individuals (e.g., households) and $t$ as time. If $f_t=1$ for all $t$, and $\lambda_i$ is scalar, then $\delta_t +\lambda_i'f_t =\delta_t+\lambda_i$, we obtain the usual additive individual and time fixed effects model. Dynamic panel models with additive fixed effects remain the workhorse for
empirical research.  The product $\lambda_i'f_t$ is known as the interactive effects \cite{Bai2009}, and is more general than additive fixed effects models. The models allow the individual heterogeneities (such as unobserved innate  ability, captured by $\lambda_i$) to have time varying impact (through $f_t$) on the outcome variable $y_{it}$. In a different perspective, the models allow common shocks (modeled by $f_t)$ to have heterogeneous impact (through $\lambda_i$) on the outcome.
For many panel data sets, $N$ is usually much larger than $T$ because it is costly to keep track of the same individuals over time. Under fixed $T$, typical estimation methods such as the least squares do not yield consistent estimation of the model parameters. Consider the special case
\begin{equation} \label{eq:fixed effects}  y_{it}= \alpha y_{it-1} + c_i  + \eps_{it} \end{equation}
where $c_i$ are fixed effects. This corresponds to $\delta_t=0$ and $f_t=1$ for all $t$.  Even if $c_i$ are iid, zero mean and finite variance, and independent of $\eps_{it}$, the least squares estimator
$\hat \alpha =(\sum_i\sum_t y_{it-1}^2)^{-1} \sum_i\sum_t y_{it-1} y_{it}$ is easily shown to be inconsistent. When $c_i$ are treated as parameters to be estimated along with $\alpha$, the least squares method is still biased, and the order of bias is $O(1/T)$, no matter how large is $N$, see \cite{Nickell1981}.  So unless $T$ goes to infinity, the least squares method remains inconsistent, an issue known as the incidental parameters problem, for example, \cite{Kiviet1995} and \cite{Lancaster2000}.

However, provided $T\ge 3$, consistent estimation of $\alpha$ is possible with the instrumental variables (IV) method. Anderson and Hsiao \cite{AndersonHsiao1982} suggested the IV estimator
by solving $ \sum_{i=1}^N y_{i1} (\Delta y_{i3}-\alpha \Delta y_{i2}) =0$, where $\Delta y_{it}=y_{it}-y_{it-1}$. Differencing the data purges $c_i$, but introduces correlation between the regressor and the resulting errors, which is why the IV method is used.  With $T$ strictly greater than 3,
more efficient IV estimator  is suggested by Arellano and Bond \cite{ArellanoBond1991}. Model (\ref{eq:fixed effects}) can also be estimated by the Gaussian quasi-maximum likelihood method, for example,
\cite{AlvarezArellano2022}, \cite{Bai2013}, and \cite{Moreira2009}.

For model (\ref{eq:dyn-interative}), differencing cannot remove the interactive effects since $ \Delta y_{it}=\alpha \, \Delta y_{it-1} +\Delta\delta_t + \lambda_i'\Delta f_t + \Delta\eps_{it}$.
The model can be estimated by the fixed effects approach, treating both $\lambda_i$ and $f_t$ as parameters. Just like the least squares for the earlier additive effects model, the fixed effects method will  produce bias. Below we introduce the quasi-likelihood approach, similar to \cite{BaiLi2012, BaiLi2014} for non-dynamic models.

Project the first observation $y_{i1}$ on $[1,\lambda_i]$ and write $ y_{i1} =\delta_1^* + \lambda_i'f_1^*  + \eps_{i1}^*$,
where $(\delta_1^*, f_1^*)$ is the projection coefficients, and $\eps_{i1}^*$ is the projection error.
The  asterisk variables are different  from the true $(\delta_1,f_1,\eps_{i1})$ that generates $y_{i1}$.  This projection is called upon because $y_{i0}$ is not observable.\footnote{The first observation starts at $t=1$, $y_{i0}$ is not available. If $y_{i0}$ were observable we would have $y_{i1}=\alpha y_{i0} +\delta_1 +\lambda_i'f_1 + \eps_{i1}$. But then a projection of $y_{i0}$ on $[1,\lambda_i]$  would be required.}
Note that  we can drop the superscript $*$ to simplify the  notation. This is because we will treat $\delta_t$ and $f_t$ as (nuisance and free) parameters, and we do not require
$\eps_{it}$ to have the same distribution over time. This means we can rewrite $y_{i1}$ as
$y_{i1}= \delta_1+ \lambda_i'f_1 + \eps_{i1}$.

The following notation will be used
\begin{equation} \label{dyn-notation}
y_i= \begin{bmatrix}
  y_{i1} \\
  \vdots \\
  y_{iT} \\
\end{bmatrix},
\quad \delta=\begin{bmatrix}
                           \delta_1 \\
                           \vdots \\
                           \delta_T \\
                         \end{bmatrix},   \quad
F=\begin{bmatrix}
      f_1' \\
      \vdots \\
      f_T' \\
    \end{bmatrix}, \quad  \eps_i= \begin{bmatrix}
                             \eps_{i1} \\
                             \vdots \\
                             \eps_{iT} \\
                           \end{bmatrix}
\end{equation}
together with the following $T\times T$ matrices,
\begin{equation} \label{JL}
 B=\begin{bmatrix}
     1 & 0 & \cdots & 0 \\
     -\alpha  & 1 & \cdots & 0 \\
     \vdots & \ddots & \ddots & \vdots\\
     0 & \cdots &  -\alpha & 1 \\
   \end{bmatrix}, \quad
  J=\begin{bmatrix}
     0 & 0 & \cdots & 0 \\
     1  & 0 & \cdots & 0 \\
     \vdots & \ddots & \ddots & \vdots\\
     0 & \cdots &  1 & 0 \\
   \end{bmatrix}, \quad  L=
   \begin{bmatrix}
     0 & 0 & \cdots & 0 & 0\\
     1  & 0 & \cdots & 0 & 0 \\
     \alpha & 1 & \ddots & 0 & 0 \\
     \vdots & \ddots & \ddots & \ddots & \vdots \\
     \alpha^{T-2} & \cdots & \alpha & 1 & 0 \\
   \end{bmatrix} \end{equation}
 Note that $L= J B^{-1}$.
With these notations, we can write the model as
\[ By_i = \delta +F \lambda_i + \eps_i \]
This gives a simultaneous equations system with $T$ equations.
We assume $\lambda_i$ are iid, independent of $\eps_i$.
Without loss of generality, we assume $\E(\lambda_i)=0$, otherwise, absorb $F \lambda$ (where $\lambda=\E\lambda_i )$ into $\delta$. Further assume
$ \E(\lambda_i\lambda_i')=I_r$ (a normalization restriction, where $I_r$ is an identity matrix).
We also assume $\eps_i$ are iid with zero mean, and
\[ D =\var(\eps_i) =\diag(\sigma_1^2, \sigma_2^2,...,\sigma_T^2)  \]
These assumptions imply that $By_i$ are iid with mean $\delta$ and covariance matrix $FF'+D$. Consider the Gaussian quasi likelihood function
\[ \ell_{NT} (\theta) =-\frac N 2 \log|FF'+D|-\frac 1 2 \sumiN (By_i-\delta)'(FF'+D)^{-1} (By_i-\delta) \]
where the Jacobian does not enter since the determinant of $B$ is 1, where $\theta=(\alpha, \delta, F, \sigma_1^2,...,\sigma_T^2)$.
The quasi-maximum likelihood estimator (QMLE) is defined as
\[ \hat \theta =\argmax_\theta \ell_{NT}(\theta) \]
The asymptotic distribution of this estimator is studied by \cite{Bai2024}.

An alternative estimator, the fixed effects estimator, treats both $\lambda_i$ and $f_t$ as parameters, in addition to $\alpha$ and $\delta_t$.  The corresponding likelihood function under normality of
$\eps_{it}$ is given in (\ref{fixedeffects-LR}) below.  The fixed effects framework estimates more nuisance parameters (can be substantially more under large $N$), the source of incidental parameters problem.  Our analysis focuses on QMLE. Comparison of the two approaches will be made.

The objectives of the present paper are threefold.
First, what is the efficiency bound for the system maximum likelihood estimator under normality assumption? Second, does the QMLE attain the normality efficiency bound with or without the normality assumption? Third, how does  QMLE fare in comparison to the fixed effects estimator?

 We approach these questions with Le Cam's type of analysis. The difficulty lies in the increasing dimension of the parameter space as
$T$ goes to infinity because the number of parameters is of order $T$.
No sparsity in parameters is assumed. With sparsity, \cite{JankovaGeer2018} derived efficiency  bounds and constructed efficient estimators
 via regularization for various models.
The  ability to deal with non-sparsity in the current model relies on panel data.

On notation: $\|A\|$ denotes the Frobenius norm for matrix (or vector) $A$, that is, $\|A\|= (\tr(A'A))^{1/2}$, and $\|A\|_2$ denotes the spectral norm of $A$, that is, the square root of the largest
eigenvalue of $A'A$. Notice $\|AB\|\le \|A\|_2 \|B\|$. The transpose of $A$ is denoted by $A'$; $|A|$ and $\tr(A)$ denote, respectively, its determinant and trace for a square matrix $A$.

\section{Assumptions for QMLE}

We assume $|\alpha|<1$ for asymptotic analysis.   The following assumptions are made for the model.

\vspace{0.1in}
\noindent
{\bf Assumption A}

(i)  $\eps_i$ are iid over $i$; $\E(\eps_{it})=0$, $\var(\eps_{it})=\sigma_t^2>0$, and $\eps_{it}$ are independent over $t$;
 $\E\eps_{it}^4 \leq M<\infty$ for all $i$ and $t$.

(ii) The $\lambda_i$ are iid,  independent of $\eps_i$, with $\E\lambda_i=0$, $\E(\lambda_i \lambda_i')=I_r$, and
$\E\|\lambda_i\|^4 \le M$.

(iii)  There exist constants
$a$ and $b$ such that $0 < a < \sigma_t^2 < b< \infty$ for all $t$;
$ \frac 1 T  F' D^{-1} F =\frac 1 T \sumtT \sigma_t^{-2} f_t f_t' \rightarrow \Sigma_{ff}>0$.

\vspace{0.1in}

Two comments are in order for this model. First, Assumption A(ii) assumes $\lambda_i$ are random variables, but they can be fixed bounded constant. All needed is that
$\Psi_N:=\frac 1 N \sumiN (\lambda_i-\bar \lambda) (\lambda_i-\bar \lambda)' \rightarrow \Psi >0$ (an $r\times r$ positive definite matrix), where $\bar \lambda$ is the  sample average of $\lambda_i$. One can  normalize the matrix
$\Psi$ to be an identity matrix.
Second, $F$ is determined up to an orthogonal rotation since $FF'=F R (FR)'$ for $RR'=I_r$. the rotational indeterminacy can be removed by the normalization that $F'D^{-1}F$  is  a diagonal matrix (with distinct elements),
see \cite{Anderson2003} (p.573) and \cite{LawleyMaxwell1971} (p.8).
Rotational indeterminacy does not affect the estimate for
$\alpha$, $D$, and $\delta$.

Under Assumption A, \cite{Bai2024} showed that the QMLE for $\hat \alpha$ has the following asymptotic representation under $N,T\rightarrow \infty$ with $N/T^3\rightarrow 0$,
\begin{equation} \label{rhohat-rho-rep}  \sqrt{NT}(\hat \alpha-\alpha) =   \bbig(\frac 1 T \tr(L\DD  L' \DD ^{-1}) \bigg)^{-1}
 \Big[
\frac 1 {\sqrt{NT}} \sumiN (L \eps_i)'\DD ^{-1}  \eps_i \Big ]+o_p(1),
\end{equation}
where $L$, $D$, and $\eps_i$ are defined earlier.
Note that $\E[(L \eps_i)'\DD ^{-1}  \eps_i] =\tr(L')=0$, and  $\E[(L \eps_i)'\DD ^{-1}  \eps_i]^2= \tr(L\DD  L' \DD ^{-1})$ with the expression
 \begin{equation*}
\frac 1 T \tr(L\DD  L' \DD ^{-1})= \frac 1 T \sum_{t=2}^T \frac 1 {\sigma_t^2}\bbig(\sigma_{t-1}^2 +\alpha^2 \sigma_{t-2}^2 + \cdots + \alpha^{2(t-2)} \sigma_1^2 \bbig). \end{equation*}
We assume the above converges to $\gamma>0$, as $T\rightarrow \infty$, that is
 \begin{equation} \label{eq:gammaNT}
\frac 1 T \tr(L\DD  L' \DD ^{-1}) \rightarrow \gamma >0.
\end{equation}
Then we have,
\[ \sqrt{NT}(\hat \alpha-\alpha) \convd \N(0, 1/\gamma). \]
 For the special case of homoskedasticity, ($\sigma_t^2=\sigma^2$ for all $t$), $\gamma=1/(1-\alpha^2)$, and hence
 $\sqrt{NT}(\hat \alpha-\alpha) \convd \N(0, 1-\alpha^2)$.

 QMLE requires no bias correction, unlike the fixed effects regression. The latter is considered by \cite{Bai2009} and \cite{MoonWeidner2017}.
 Our objective is to show that $1/\gamma$ is the efficiency bound under normality assumption, and  QMLE  attains the normality efficiency bound.
 This result is obtained in  the presence of increasing number of incidental parameters. The estimator $\hat \alpha$  is also consistent under fixed $T$ in contrast to
 the fixed effects estimator.  The  estimated $ f_t, \sigma_t^2$ are all $\sqrt{N}$ consistent and asymptotically normal. In particular, the estimated factors $\hat f_t$ have the asymptotic representation, for each  $t=1,2,...,T$
 \begin{equation} \label{eq:fhat-f} \sqrt{N} (\hat f_t -f_t) = \frac 1 {\sqrt{N}} \sumiN \lambda_i \eps_{it} +o_p(1)   \end{equation}
 This implies $\sqrt{N} (\hat f_t -f_t)\convd \N(0,\sigma_t^2 I_r)$.
 Details are given in Bai \cite{Bai2024}. Also see \cite{BaiLi2012} for non-dynamic factor models.

\section{Local likelihood ratios and efficiency bound}

\subsection{Related literature}

A closely related work is that of Iwakura and Okui \cite{IwakuraOkui2014}.
They consider the fixed effects framework instead of the QMLE.
The fixed effects estimation procedure treats both $\lambda_i$ and $f_t$ as parameters $(i=1,2,...,N; t=1,2,...,T$), along with $\alpha$ and $\delta$.
The corresponding likelihood function under normality of $\eps_{it}$ is\footnote{The fixed effects likelihood does not have a global maximum under heteroskedasticity, for example,  \cite{Anderson2003} (p.587), but local maximization is still meaningful.
Another solution is to impose homoskedasticity.}
\begin{equation}\label{fixedeffects-LR} \ell_{\text{fixed effects}}(\theta)=-\frac N 2 \sum_{t=2}^T  \log \sigma_t^2
-\frac 1 2 \sum_{i=1}^N \sum_{t=2}^T(y_{it}-\alpha y_{i,t-1}-\delta_t-\lambda_i'f_t)^2/\sigma_t^2  \end{equation}
The fixed effects estimator for $\alpha$ will generate bias, similar to the  fixed effects estimator for dynamic panels with additive effects. The bias is studied by \cite{MoonWeidner2017}.\footnote{
In contrast, the QMLE does not generate bias under fixed $T$.}  Iwakura and Okui \cite{IwakuraOkui2014} derive the efficiency bound for the fixed effects estimators under homoskedasticity ($\sigma_t^2=\sigma^2$ for all $t$).
Another closely related work is that of \HK \cite{HahnKuersteiner2002}. They consider the efficiency bound problem under the fixed effects framework for the additive effects model described by (\ref{eq:fixed effects})  without a factor structure. Throughout this paper, the fixed effects framework refers to methods that also estimate the factor loadings
$\lambda_i$ in addition to $f_t$.

 In contrast, we consider the likelihood function for the system of equations
 \begin{equation} \label{eq:qmle2} -\frac N 2 \log|FF'+D|-\frac 1 2 \sumiN (By_i-\delta)'(FF'+D)^{-1} (By_i-\delta) \end{equation}
 QMLE does not estimate $\lambda_i$ (even if they are fixed constants as explained earlier), thus eliminating the incidental parameters in the cross-section dimension.
The incidental parameters are now $\delta$, $F$ and $D$, and the number of parameters increases with $T$. Despite fewer number of incidental parameters,
the analysis of local likelihood  is more  demanding than that of the fixed effects likelihood (\ref{fixedeffects-LR}). Intuitively, the fixed effects likelihood (\ref{fixedeffects-LR}) is quadratic in $F$, but the QMLE likelihood $\ell(\theta)$ in (\ref{eq:qmle2}) depends on $F$ through the inverse matrix $(FF'+D)^{-1}$ and through the log-determinant of this matrix.
The high degree of nonlinearity makes the perturbation analysis more challenging.
As  demonstrated later, the local analysis brings insights regarding the relative merits of the QMLE and the fixed effects estimators.

Notice  $By_i-\delta = B(y_i-\delta^\dag)$, where $\delta^\dag=B^{-1}\delta$ is a vector of free parameters because  $\delta$ is a vector of free parameters. The concentrated likelihood function by concentrating out $\delta^\dag$ (its maximum likelihood estimator is simply $\bar y =\frac 1 N \sumiN y_i$) is given by
\begin{equation} \label{eq:concen-qmle} \ell(\theta) =-\frac N 2 \log|FF'+D|-\frac 1 2 \sumiN (y_i-\bar y)'B'(FF'+D)^{-1} B(y_i-\bar y) \end{equation}
This is the likelihood function studied by \cite{Bai2024} and \cite{BaiMones2025} for the QMLE, with the latter paper focusing on global identification. Our efficiency bound analysis is based on (\ref{eq:concen-qmle}).

There is a substantial body of research on dynamic models with interactive effects. For example, \cite{BaiLi2021} and \cite{shilee2017} study spatial models, \cite{miaophillipssu2023} examine high-dimensional vector autoregressions, and \cite{lamyao2012} focus on high-dimensional time series models. However, unlike the present work, these studies do not address efficiency bounds. Throughout this paper, we assume that the number of factors is known. In practice, the number of factors can be estimated using methods such as the one proposed by \cite{lamyao2012}.

\subsection{The \texorpdfstring{$\ell^\infty$}{Lg} local parameter space}
 Local likelihood ratio processes are indexed by local parameters.
Since the convergence rate for the estimated parameter of $\alpha^0$ is $\sqrt{NT}$, that is,
$\sqrt{NT} (\hat \alpha-\alpha^0) =O_p(1)$,
it is natural to consider the local parameters  of the form
\[ \alpha^0+ \rootNT \tilde \alpha \]
where $\ta \in \R$. However, the consideration of local parameters for $f_t^0$ is non-trivial, as explained by Iwakura and Okui \cite{IwakuraOkui2014} for the fixed effects likelihood ratio.
We consider the following local parameters
\begin{equation} \label{eq:dev1}  f_t^0+ \rootN ( \rootT \tilde f_t ), \quad  t=1,2,...,T \end{equation}
where $\|\tilde f_t\| \le M <\infty$ for all $t$ with $M$  arbitrarily given;  $\|\cdot\|$ denotes the r-dimensional Euclidean  norm. In view that the estimated factor $\hat f_t$ is $\sqrt{N}$ consistent,
that is, $\sqrt{N}(\hat f_t-f_t^0)=O_p(1)$, one would expect local parameters in the form $f_t^0+ N^{-1/2} \tf_t$,  the extra scale factor $T^{-1/2}$ in the above local rate looks rather unusual.  However, (\ref{eq:dev1}) is the suitable local rate for the local likelihood ratio to be $O_p(1)$, as is shown in both the statement and the proof of  Theorem \ref{thm:localLR-1} below.   Without the scale factor $T^{-1/2}$, the local likelihood  ratio will diverge to infinity (in absolute values) if no restrictions are imposed on $\tf_t$ other than its boundedness.  This type of local parameters was used in earlier work by \cite{HahnKuersteiner2002} for additive fixed effects estimator. Later we shall consider a different type of local parameters without the extra scale factor  $T^{-1/2}$, but other restrictions   on $\tf_t$ will be needed.

Additionally, even if one regards $\rootNT \tf_t$ to be small (relative to $\rootN \tf_t$), it is for the better provided that the associated efficiency bound is achievable by an estimator. This is because the smaller the perturbation, the lower the efficiency bound,   and hence harder to attain by any estimator.

Consider the space
\[ \ell_r^\infty: =\{ (\tilde f_t)_{t=1}^\infty  \,  \Big| \tilde f_t \in \R^r, \,  \sup_s \|\tilde f_s\| <\infty \}\]
the space of bounded sequences, each coordinate is $\R^r$-valued. Let
\[ \tf =(\tf_1,\tf_2, ...) \in \ell_r^\infty \]
and define $\tilde F=(\tf_1,\tf_2,...,\tf_T)'$,
the projection of $\tf$ onto the first $T$ coordinates.  The matrix  $\tilde F$ is $T\times r$, but we suppress its dependence on $T$ for  notational simplicity. Since $\tf \in \ell_r^\infty$, it follows that

\begin{equation} \label{eq:tFtF/T} \frac 1 T \tF'\tF =\frac 1 T \sum_{t=1}^T \tf_t \tf_t'=O(1) \end{equation}

To simplify the analysis, we assume that $D$ is known. This assumption does not affect the efficiency bound for $\alpha$, but it does simplify the derivation considerably. Nonetheless, it may be worthwhile to rigorously analyze the case where $D$ is unknown. In general, treating some parameters as known can result in a lower efficiency bound, which may be difficult to attain. In our setting, however, we demonstrate that the efficiency bounds are achievable.

Let $\theta^0=(\alpha^0,  F^0)$ and $\tilde \theta=(\ta,  \tF)$,  we study the asymptotic behavior of
\[ \ell(\theta^0+\rootNT \tilde \theta) -\ell(\theta^0) \]
under the normality of $\eps_{it}$ and $\lambda_i$. The normality assumption allows us to derive the parametric efficiency bound in the presence of increasing number
of nuisance parameters. We then show the QMLE without normality attains the efficiency bound.
In the rest of the paper, we use
$(\alpha^0,  F^0)$ and $(\alpha,  F)$ interchangeably (they represent the true parameters); $(\ta,  \tF)$ represent local parameters, and $\hat \alpha$  and  $\hat F=(\hat f_1,...,\hat f_T)'$ are the QMLE estimated parameters.

\vspace{0.1in}
\noindent
{\bf Assumption B}

(i):  $\eps_{it}$ are iid over $i$, and independent over $t$ such that $\eps_{it}\sim \N(0,\sigma_t^2)$.

(ii): $\lambda_i$ are iid $\N(0,I_r)$, independent of $\eps_{it}$ for all $i$ and $t$.

(iii):  $\sigma_t^2 \in [a, b]$ with  $ 0< a < b < \infty$ for all $t$.

(iv): $\|f_t\|\le M<\infty$ for all $t$, and $\frac 1 T F'D^{-1}F\rightarrow \Sigma_{ff}>0$, where $D=\diag(\sigma_1^2,\sigma_2^2,...,\sigma_T^2)$.

(v): As $T\rightarrow \infty$,\\
\indent \indent (a) $\frac 1 T \tr(L'D^{-1} L D)  \rightarrow \gamma >0$,\\
\indent \indent (b) $\frac 1 T \tr[ (LF)' (\DMDFD  )(LF)]  \rightarrow \nu \ge 0$\\
\indent \indent where $M_{D^{-1/2}F}$ denote the projection matrix orthogonal to $D^{-1/2} F$. Specifically,
\[ \DMDFD =D^{-1}-D^{-1}F(F'D^{-1}F)^{-1}F'D^{-1}. \]

\vspace{0.1in}

Under assumptions B(i) and (ii), $F \lambda_i +\eps_i$   are iid $\N(0, FF'+D)$, implying a parametric model with an increasing dimension of incidental parameters.
Normality  for $\lambda_i$ and $\eps_{it}$  is a standard assumption in factor analysis, see, e.g., Anderson \cite{Anderson2003} (p.576). Here we switch the role of $\lambda_i$ and $f_t$. Note in classical factor analysis, the time dimension $T$ (in our notation) is fixed, there is no incidental parameters problem since the number of parameters is fixed. But we consider $T$ that goes to infinity.
The following theorem gives the asymptotic representation for the local likelihood ratios.

\begin{theorem} \label{thm:localLR-1} Under Assumption B,  for $\ta\in \R$, $\tf\in \ell_r^\infty$,  $\tF=(\tf_1,\tf_2,...,\tf_T)'$,
 we have
as $N,T\rightarrow \infty$, with $N/T^3\rightarrow 0$,
\[ \ell(\theta^0+\rootNT \tilde \theta) -\ell(\theta^0) =\Delta_{NT}(\tilde \theta)-\frac 1 2 \E[\Delta_{NT}(\tilde \theta)]^2 + \op\]
where
\begin{align}
  \Delta_{NT}(\tilde \theta) & = \rootNT \sumiN \lambda_i'\tF'[\DMDFD ]\eps_i  \notag\\
    & + \talpha \, \rootNT \sumiN (L\eps_i)'D^{-1}\eps_i   \label{eq:DeltaNT}\\
  & +\talpha \, \rootNT \sumiN\lambda_i' (LF)'[\DMDFD ] \eps_i  \notag
  \end{align}
\begin{align}
 \E[\Delta_{NT}(\tilde \theta)]^2= & \frac 1 T \tr \Big[\tF'[\DMDFD] \tF \Big] \notag \\
  & + \talpha^2  \Big[ \frac 1 T \tr(L'D^{-1}LD ) \Big] \label{eq:varDeltaNT}\\
  & + \talpha^2  \tr \Big[ \frac 1 T  (LF)' [\DMDFD  ](LF)\Big] \notag\\
    & + 2\talpha \,  \tr \Big[ \frac 1 T (LF)'[\DMDFD ] \tF \Big] \notag
       \end{align}
where $o_p(1)$ is uniform  over $\tilde \theta$ such that $|\ta|\le M$,
$ \frac 1 T \|\tF'\tF \|=\frac 1 T \sum_{t=1}^T \|\tf_t\|^2 \le M$,   for any given $M<\infty$.
\end{theorem}

The proof of Theorem \ref{thm:localLR-1} is given in the \hyperref[appn]{Appendix}.

Note that the expected value of $\Delta_{NT}(\tilde \theta)$ is zero,   so $\E[\Delta_{NT}(\tilde \theta)]^2$ is the variance.

All terms in $\Delta_{NT}(\ttheta)$ are stochastically bounded, they have expressions of the form $\rootNT \sumiN\sumtT \xi_{it}$, where
$\xi_{it}$ have zero mean, and finite variance (and in fact finite moments of any order
under Assumption B).
Assume $\tf$ satisfies
\begin{align*}  \frac 1 T \tF' D^{-1} \tF & =\frac 1 T \sumtT \frac 1 {\sigma_t^2}  \tf_t \tf_t' \rightarrow \Sigma_{\tf \tf}
\end{align*}
as well as existence of limits for $\frac 1 T \tF'D^{-1} F$, then
the variance $\E [\Delta_{NT}(\ttheta)]^2$ has a limit. Let
$\E [\Delta_{NT}(\ttheta)]^2 \rightarrow \tau^2$ for some $\tau^2$ depending on $(\ta, \tf)$.
We can further show
\[ \Delta_{NT}(\tilde \theta) \convd \N(0, \tau^2). \]
Thus the local likelihood ratio can be rewritten as
\[  \ell(\theta^0+\rootNT \tilde \theta) -\ell(\theta^0) =\Delta_{NT}(\tilde \theta)-\frac 1 2 \tau^2  + \op.\]

We next consider the asymptotic efficiency bound for regular estimators. Regular estimators rule out  Hodges type ``superefficient'' and James-Stein type estimators. A regular estimator sequence converges locally uniformly (under the local laws)  to a  limiting distribution that is free from the local parameters (van der Vaart \cite{vanderVaart1998}, p.115, p.365).

\subsection{Efficient scores and efficiency bound}

 In the likelihood ratio expansion, the  term $\Delta_{NT}(\tilde \theta)$ contains the scores of the likelihood function. The coefficient of
$\ta$ gives the score for $\alpha^0$, and the coefficient of $\tf_t$ gives the score of $f_t^0$.  The efficient score for $\alpha^0$ is the projection residual of its own score onto the scores of $f_1^0,...,f_T^0$.  Moreover, the inverse of the variance of the efficient score gives the efficiency bound (Bickel et al, \cite{bickel1993}, p. 28).

To derive the efficient score for $\alpha^0$, rewrite $\Delta_{NT}(\tilde \theta)$ of Theorem 1 as
\[ \Delta_{NT}(\tilde \theta)=\Delta_{NT1} +  \talpha\, [ \Delta_{NT2}+\Delta_{NT3}]  \]
where  $\Delta_{NT1}$  denotes the first term of $\Delta_{NT}(\tilde \theta)$, see (\ref{eq:DeltaNT}), and
$\Delta_{NTj}$ $(j=2,3$) denote the last two terms of (\ref{eq:DeltaNT}), but taking out $\ta$.
So the score for  $\alpha^0$
is the sum $\Delta_{NT2}+\Delta_{NT3}$.  Next, rewrite
\begin{align*} \Delta_{NT1} & = \rootNT \sumiN \lambda_i'\tF'[\DMDFD ]\eps_i =\frac 1 {\sqrt{T}} \sum_{t=1}^T \tf_t'  v_t
 \end{align*}
where  $v_t= \rootN \sumiN \lambda_i v_{it} $ $(r\times 1)$ and $v_{it}$ is the $t$-th element of the vector $[\DMDFD ]\eps_i$.
 Thus $v_t$ is the score  of $f^0_t$  ($t=1,2,...,T$).
 To obtain the efficient score for $\alpha^0$, we project $\Delta_{NT2}+\Delta_{NT3}$ onto the scores of $f_t^0$, that is onto
 $[v_1,v_2....,v_T]$to get the projection residuals.  Let $V_T=(v_1',v_2',...,v_T')'$. The projection residual is given by
\begin{equation}\label{eq:proj-resid}   \Delta_{NT2}+\Delta_{NT3} -V_T'[ \E(V_T V_T')]^{-1} \E \left[V_T (\Delta_{NT2} +\Delta_{NT3})\right]
\end{equation}
Notice $\Delta_{NT2}$ is uncorrelated with the scores of $f_t^0$, i.e.  $\E(V_T \Delta_{NT2})=0$. This follows because $L\eps_i$ contains the lags of
$\eps_i$, so $\Delta_{NT2}$ is composed of  terms  $\eps_{it-s}\eps_{it}$ (with $s\ge 1$), and $\E (\eps_{it-s}\eps_{it} \eps_{ik})=0$ for any $k$.
 Next, $\Delta_{NT3}$ is simply a linear combination of $V_T= [v_1,v_2,...,v_T]$ since
 $\Delta_{NT3}$ can be written as $T^{-1/2} \sum_{t=1}^T p_t' v_t$, where $p_t'$ is the $t$-th row of the matrix $L F$.
 Thus $V_T'[ \E(V_T V_T')]^{-1} \E [V_T (\Delta_{NT3})]\equiv \Delta_{NT3}$.
  In summary, we have
  \[ V_T'[ \E(V_T V_T')]^{-1} \E \left[V_T (\Delta_{NT2} +\Delta_{NT3}\right] =\Delta_{NT3} \]
  It follows that the projection residual in (\ref{eq:proj-resid}) is equal to $\Delta_{NT2}$.
 Hence the efficient score for $\alpha^0$  is  $\Delta_{NT2}$.
   Notice,
 \begin{equation} \label{eq:gamma} \var(\Delta_{NT2})= \frac 1 T \tr(L'D^{-1} L D)= \frac 1 T \sum_{t=2}^T \frac 1 {\sigma_t^2}\bbig(\sigma_{t-1}^2 +\alpha^2 \sigma_{t-2}^2 + \cdots + \alpha^{2(t-2)} \sigma_1^2 \bbig)   \end{equation}
  its limit is $\gamma$ by Assumption B(v), so $1/\gamma$ gives the asymptotic  efficiency bound.

  We summarize the result in the following corollary.
\begin{corollary} \label{coro:bound1} Under Assumption B, the asymptotic efficiency bound for regular estimators of $\alpha^0$,
is $1/\gamma$, with $\gamma$ being the limit of (\ref{eq:gamma}).
\end{corollary}

Since Assumption B is stronger than Assumption A,
 the asymptotic representation in (\ref{rhohat-rho-rep}) holds under Assumption B. That is,  under the normality assumption, the system maximum likelihood
 estimator satisfies
\[   \sqrt{NT}(\hat \alpha-\alpha) =   \bbig(\frac 1 T \tr(L\DD  L' \DD ^{-1}) \bigg)^{-1}
 \Big[
\frac 1 {\sqrt{NT}} \sumiN (L \eps_i)'\DD ^{-1}  \eps_i \Big ]+o_p(1).  \]
  We see that $\sqrt{NT}(\hat \alpha-\alpha^0)$ is expressed in terms of
the efficient influence functions, thus $\hat \alpha$  is regular and  asymptotically efficient (van der Vaart \cite{vanderVaart1998}, p.121 and p.369). We state the result in the following corollary.

\begin{corollary}  Under Assumption B, the system maximum likelihood estimator  $\hat \alpha$ is a regular estimator and achieves the asymptotic efficiency bound (in spite  of an increasing number of incidental parameters).
\end{corollary}

The preceding corollaries imply that, under normality, we are able to establish the asymptotic efficiency bound in the presence of increasing number of
nuisance parameters. Further, the system maximum likelihood  estimator achieves the efficiency bound.  These results are not obvious owing to the incidental parameters problem.

QMLE in Section 2 does not require normality.  But it achieves the normality efficiency bound, see equation (\ref{rhohat-rho-rep}).
So QMLE is robust to the normality assumption.
If $\lambda_i$ and $\eps_i$ are non-normal, and their distributions are known, one should be able to construct a more efficient estimator than the QMLE.
But for panel data analysis in practice, researchers usually do not impose distributional assumptions other than some moment conditions such as those in Assumption A.
Thus QMLE presents a viable estimation procedure, knowing that it achieves the normality efficiency bound. Furthermore, QMLE does not need bias correction, unlike the fixed effects estimator.

\subsection{Constructing a Hilbert subspace}

The result of Corollary 1 is not directly obtained via a limit experiment and the convolution theorem  (e.g.,  van der Vaart and Wellner, \cite{vanderVaartWellner1996}, chapter 3.11).
Since $\ell_r^\infty$ is not a Hilbert space,  the convolution theorem  is not directly applicable. However, using the line of argument
in \cite{IwakuraOkui2014} it is possible to construct a Hilbert subspace with an appropriate inner product  in which the efficiency bound for the low dimensional parameter $\alpha^0$ can be shown to be  $1/\gamma$. That is, Corollary 1 can be obtained via the convolution theorem.


We now construct a Hilbert subspace so that convolution theory can be directly applied. For this purpose, we adopt the method proposed by \cite{IwakuraOkui2014}.

\vspace{0.1in}
\noindent
{\bf Assumption C.}
Let $\psi$ be a vector of continuous function from [0,1] to $\R^r$, and $f_t =\psi(t/T)$. Let $\sigma$ be a continuous function from [0,1] to $\R$ and $b\ge\sigma(s)\ge a>0$ for all $s \in [0,1]$, and $\sigma_t^2 =\sigma^2(t/T)$.  They satisfy $ \frac 1 T \sumtT \frac 1 {\sigma_t^2} f_t f_t'\rightarrow \int_0^1 \frac 1 {\sigma(s)^2} \psi(s) \psi(s)' ds >0$.
\vspace{0.1in}

Let $C([0,1]; \mathbb{R}^r)$ be the space of continuous \(\mathbb{R}^r\)-valued functions on the interval \([0,1]\).
That is,
\[
C([0,1]; \mathbb{R}^r) = \left\{ g : [0,1] \to \mathbb{R}^r \mid g \text{ is continuous} \right\}
\]
Let $C_f$ be the subspace of $C([0,1]; \mathbb{R}^r)$ that is  orthogonal to $\psi$ weighted by  $\sigma \in C[0,1]$. That is, for each $\tilde \psi \in C_f$, we have
$\int_0^1  \frac 1 {\sigma(s)^2} \tilde \psi(s) \psi'(s) ds =0$. Clearly, $C_f$ is a linear space. If we define $\langle g_1, g_2\rangle
 =\int_0^1 \frac 1 {\sigma(s)^2} g_1(s)' g_2(s) ds$ for any $g_1,g_2\in C_f$, then $\langle \cdot, \cdot\rangle$ forms an inner product.
In particular, $\langle g, g\rangle =0$ if and only if $g=0$.

We consider the local parameter space
\[  \H= \R \times C_f \]
For any $h_1=(\tilde \alpha_1, \tilde \psi_1)\in \H$ and $h_2=(\tilde \alpha_2, \tilde \psi_2)\in \H$, define
\[ \langle h_1, h_2\rangle_{\H}= \ta_1 \ta_2 \gamma + \int_0^1 \frac 1 {\sigma(s)^2} \tpsi_1(s)' \tpsi_2(s) ds \]
where $\gamma>0$ is given in (\ref{eq:gammaNT}),
then $\langle \cdot,\cdot\rangle_{\H}$ is an inner product on $\H$, and equipped with it, $\H$ becomes a Hilbert space.

For each $\tpsi \in C_f$, we define $\tilde f_t =\tpsi(t/T)$, and the local parameter
 $f_t^0 + \tilde f_t/\sqrt{NT}$, $t=1,2,...,T$. Let $\tilde F =(\tilde f_1,\tilde f_2,...,\tilde f_T)'$ be a $T\times r$ matrix.

\begin{theorem} \label{thm:localLR-1new} Under Assumptions Theorem 1 and Assumption C,  we have
\[ \ell(\theta^0+\rootNT \tilde \theta) -\ell(\theta^0) =\Delta_{NT}(\tilde \theta)-\frac 1 2 \E[\Delta_{NT}(\tilde \theta)]^2 + \op\]
where
\begin{align}
  \Delta_{NT}(\tilde \theta) & = \rootNT \sumiN \lambda_i'\tF' D^{-1}\eps_i
    + \talpha \, \rootNT \sumiN (L\eps_i)'D^{-1}\eps_i   \label{eq:DeltaNT2}
 \end{align}
\begin{align*}
 \E[\Delta_{NT}(\tilde \theta)]^2= & \frac 1 T \tr \Big[\tF'D^{-1} \tF \Big]
   + \talpha^2  \Big[ \frac 1 T \tr(L'D^{-1}LD ) \Big]
  \end{align*}
where $o_p(1)$ is uniform  over $\tilde \theta$ such that $|\ta|\le M$,
$ \frac 1 T \|\tF'\tF \|=\frac 1 T \sum_{t=1}^T \|\tf_t\|^2 \le M$,  for any given $M<\infty$.
\end{theorem}

The proof of Theorem \ref{thm:localLR-1new} is in the Appendix.
From
\[ \frac 1 T \tr (\tF' D^{-1}\tF) = \frac 1 T \sum_{t=1}^T \frac 1 {\sigma(t/T)^2} \tpsi(t/T)'\tpsi(t/T) \rightarrow
\int_0^1 \frac 1 {\sigma(s)^2} \tpsi(s)' \tpsi(s) ds \]
and by (\ref{eq:gammaNT}), $\frac 1 T \tr(L'D^{-1}LD )\rightarrow \gamma>0$, we have

\[ \E[\Delta_{NT}(\tilde \theta)]^2 \longrightarrow \ta^2 \gamma + \int_0^1 \frac 1 {\sigma(s)^2} \tpsi(s)' \tpsi(s) ds \]
The right hand side is simply $\|h\|_{\H}^2$ for $h=(\ta, \tpsi) \in \H$. That is,
\[ \|h\|^2_{\H} =\langle h, h\rangle= \ta^2 \gamma + \int_0^1 \frac 1 {\sigma(s)^2} \tpsi(s)' \tpsi(s) ds \]
We can write
\[ \E[\Delta_{NT}(\tilde \theta)]^2 = \| h\|^2_{\H} +o(1). \]

Also note that $\Delta_{NT}(\tilde \theta)$ consists of two terms, and each of them is asymptotically normal, and the two terms are also asymptotically independent, thus
\[ \Delta_{NT}(\tilde \theta) \convd N(0, \|h\|_{\H}^2). \]
Summarizing results, we have
\begin{corollary} Under Assumptions B and C, we have
\[ \ell(\theta^0+\rootNT \tilde \theta) -\ell(\theta^0) = \Delta_{NT}(\tilde \theta) -\frac 1 2 \|h\|_{\H}^2 + o_p(1) \]
and  $\Delta_{NT}(\tilde \theta) \convd N(0, \|h\|_{\H}^2).$
\end{corollary}

It is also straightforward to show that the likelihood ratio process is  LAN (locally asymptotically normal).

The convolution theorem  (\cite{vanderVaartWellner1996}, chapter 3.11) implies that the efficiency bound for any regular estimators of $\alpha^0$ is $1/\gamma$,
which is the inverse of the coefficients of $\ta^2$ in $\|h\|_{\H}^2$ (see the detailed argument for the application of the convolution theorem following Corollary \ref{coro:coro4} below). The efficiency bound coincides with the result in Corollary \ref{coro:bound1}.

As noted  by \cite{IwakuraOkui2014}, however, the parameter sequence $f_t^0+\tf_t/\sqrt{N}$ is not regular, the convolution theory is not applicable for deriving the efficiency bound for regular  estimators of $f_t^0$. The underlying reason for the non-regularity is that the linear functional, say, $\dot \kappa: C_f \rightarrow \R$
such that $\dot \kappa(\tpsi) =\tpsi(x_0)$, for an $x_0\in [0,1]$ is not a continuous functional. The non-continuity is because convergence in the $C_f$ space with the given norm  does not imply  pointwise convergence.
This motivates us to consider the $\ell^2$ norm in the next section, in the hope that the convolution theorem will be applicable for all parameters, including the incidental ones.

\vspace {0.1in}
\subsection{The \texorpdfstring{$\ell^2$}{Lg} local parameter space}

The previous result is only applicable for low dimensional parameters.
To apply the convolution theorem
(\cite{vanderVaartWellner1996}, chapter 3.11) to all parameters including the incidental ones,  we consider the  second type of local parameter space,
which is also used by \cite{IwakuraOkui2014} for the fixed effects estimators:
\begin{equation} \label{eq:dev2}  f_t^0 +\rootN \tf_t, \quad t=1,2,...,T\end{equation}
with $\tf=(\tf_1,\tf_2,...)$ being required to be in $\ell_r^2$:
\[ \ \ell_r^2 := \Big\{ (\tf_t)_{t=1}^\infty \,  \Big| \tf_t \in \R^r, \sum_{s=1}^\infty \|\tf_s\|^2 <\infty \Big\}. \]
For this type of local parameters,  we can remove the scale factor $T^{-1/2}$, (cf. (\ref{eq:dev1})). Since
 $\tf \in \ell_r^2$, we have,  for $\tF=(\tf_1,\tf_2,...,\tf_T)'$ (projection of  $\tf$ on the first $T$ coordinates),
 \[ \tF'\tF = \sum_{t=1}^T \tf_t \tf_t' = O(1). \]

\begin{theorem} \label{thm:localLR-2} Under Assumption B, for $\ta \in \R$,
$\tf\in \ell_r^2$,  $\tF=(\tf_1,\tf_2,...,\tf_T)'$,
as $N,T\rightarrow \infty$, with $N/T^3\rightarrow 0$, we have
\[ \ell(\alpha^0+\frac {\ta} {\sqrt{NT}},   F^0+\rootN \tF) -\ell(\theta^0)
 =\Delta_{NT}^\dag(\tilde \theta)-\frac 1 2 \E[\Delta_{NT}^\dag(\tilde \theta)]^2 + \op\]
where
\begin{align}
  \Delta_{NT}^\dag(\tilde \theta) & = \rootN \sumiN \lambda_i'\tF'D^{-1}\eps_i  \notag \\
  & + \talpha \, \rootNT \sumiN (L\eps_i)'D^{-1}\eps_i  \label{eq:Delta*}  \\
  & +\talpha \, \rootNT \sumiN\lambda_i' (LF)'[\DMDFD] \eps_i \notag  \\
 \E[\Delta_{NT}^\dag(\tilde \theta)]^2 & = \tr \Big[\tF'D^{-1} \tF \Big]  \notag \\
  & + \talpha^2  \Big[ \frac 1 T \tr(L'D^{-1}LD ) \Big]  \label{eq:variance-Delta*}\\
  & + \talpha^2  \tr \Big[ \frac 1 T  (LF)' [\DMDFD  ](LF)\Big] \notag
     \end{align}
 where $o_p(1)$ is uniform over $|\ta|\le M$,  and $\|\tF\|\le M$ for any given $M<\infty$.
\end{theorem}

In comparison with Theorem \ref{thm:localLR-1}, Theorem  \ref{thm:localLR-2} has simpler expressions, due to the smaller local parameter space.
The first two terms in $\Delta_{NT}(\tilde \theta)$ are simplified, with the corresponding simplification
in the variance, and in addition, the covariance term  in $\E[\Delta_{NT}(\tilde \theta)]^2$ is dropped. The proof of Theorem \ref{thm:localLR-2} is given in the appendix.

We next establish the local asymptotic normality (LAN) property for the local likelihood ratios.

For each $(\ta,\tf)=(\ta, \tf_1,\tf_2,....) \in  \R\times \ell_{r}^2$, we introduce a new sequence
\begin{equation}\label{eq:h} h(\ta,\tf)=(h_0,h_1,h_2,...) =\Big(\ta (\gamma+\nu)^{1/2}, \frac 1 {\sigma_1}{\tf_1}, \frac 1  {\sigma_2} {\tf_2},...\Big) \end{equation}
so $h_0= \ta (\gamma+\nu)^{1/2}$, and $h_s= \frac 1 {\sigma_s} \tf_s$ for $s\ge 1$, where $\gamma$, $\nu$,  are defined in Assumption B(v), and $\sigma_s^2$ is the variance  of $\eps_{is}$.   Hence, $h(\ta,\tf)$ is a scaled version of $(\ta,\tf)$. By Assumption B(iii),
$\min_s \sigma_s^2\ge a >0$, it follows that $h(\ta,\tf) \in \H := \R\times \ell_{r}^2$. For any $h,g\in \H$, define the inner product,
 $\langle h,g\rangle = h_0 g_0 + \sum_{s=1}^\infty h_s'g_s$, then $\H$ is a Hilbert space. Let  $\|h\|_{\H}^2=\langle h, h\rangle$. In particular,
 for $h=h(\ta,\tf)$ in (\ref{eq:h}), we have
 \begin{equation} \label{eq:hnorm} \|h\|_{\H} ^2 = \ta^2 (\gamma+\nu)  + \sum_{s=1}^\infty  \frac 1 {\sigma_s^2} \tf_s'\tf_s  \end{equation}

Notice
$ \tr(\tF' D^{-1} \tF)=\sum_{s=1}^T\frac 1 {\sigma_s^2} \tf_s'\tf_s = \sum_{s=1}^\infty  \frac 1 {\sigma_s^2} \tf_s'\tf_s +o(1)$ because the series is convergent.  By Assumption B(v), we can write
(\ref{eq:variance-Delta*}) as
\begin{equation} \label{eq:normH} \E[\Delta_{NT}^\dag(\tilde \theta)]^2 = \ta^2 (\gamma+\nu)  + \sum_{s=1}^\infty  \frac 1 {\sigma_s^2} \tf_s'\tf_s +o(1) =\|h\|_{\H} ^2 +o(1) \end{equation}
where $h =h(\ta,\tf)$ is given in (\ref{eq:h}).
Next, rewrite (\ref{eq:Delta*}) as
\[ \Delta_{NT}^\dag(\tilde \theta) =  \rootN \sumiN \lambda_i'\tF'D^{-1}\eps_i  +\ta ( \Delta_{NT2}+\Delta_{NT3}) \]
 where $\Delta_{NTj}$ ($j=2,3)$  are defined
earlier. The first term
\[  \rootN \sumiN \lambda_i'\tF'D^{-1}\eps_i =\sumtT \frac 1 {\sigma_t^2} \tf_t' \Big(\rootN \sumiN \lambda_i \eps_{it}\Big)
\convd \N\Big(0, \sum_{t=1}^\infty \tf_t'\tf_t/\sigma_t^2 \Big) \]
because $N^{-1/2}\sumiN \lambda_i \eps_{it} \convd \N(0,I_r)$.
The LHS above is asymptotically independent of  $\Delta_{NT2}+\Delta_{NT3}$ (their covariance being zero, as is shown in the appendix). From
\[ \Delta_{NT2}+\Delta_{NT3}\convd \N(0, \gamma+\nu ) \]
where $\gamma$ and $\nu$ are given in Assumption B,
we have
\[ \Delta_{NT}^\dag(\tilde \theta) \convd \N\left(0, \|h\|_{\H}^2 \, \right). \]

Moreover, it is not difficult to establish the finite dimensional weak convergence. Let $\ta^{j}\in \R$, $\tf^{j}\in \ell_{r}^2$ for $j=1,2,..,q$. Let $h^j=h(\ta^j,\tf^{j})$ and  $\tilde\theta^j=(\ta^j, \tF^{j})$, for any finite
integer $q$,
\[ (\Delta_{NT}^\dag(\tilde \theta^1),...,\Delta_{NT}^\dag(\tilde \theta^q))' \convd  \N\Big(0, ( \langle h^j, h^k\rangle)_{j,k=1}^q \Big). \]
Summarizing the above, we have
\begin{corollary} \label{coro:coro4} Under the assumption of Theorem \ref{thm:localLR-2},
\[ \ell(\alpha^0+\frac {\ta} {\sqrt{NT}}, F^0+\rootN \tF) -\ell(\theta^0) =\Delta_{NT}^\dag(\tilde \theta)  -\frac 1 2 \|h\|_{\H}^2 +o_p(1) \]
and
\[ \Delta_{NT}^\dag(\tilde \theta)\convd \N\left(0, \|h\|_{\H}^2\right) \]
where $h=h(\ta,\tf)$ and $\|h\|_{\H}^2$ are defined in (\ref{eq:h}) and (\ref{eq:hnorm}), respectively. 
Furthermore,  the likelihood ratio is  locally asymptotically normal (LAN).
\end{corollary}

Using the convolution theorem for locally asymptotically normal (LAN) experiments,  the implied efficiency bound for regular estimators of $\alpha^0$ is $1/(\gamma+\nu)$.  The implied efficiency bound for
regular estimators of $f_t^{0}$  is $\sigma_t^2 I_{r}$ for each $t$. These bounds are, respectively,  the inverse of the coefficient of $\ta^2$, and the inverse of the  matrix in  the quadratic
form  $\tf_t'\tf_t/\sigma_t^2 = \tf_t'(I_{r}/\sigma_t^2) \tf_t$  in the expression for  $\|h\|_{\H}^2$.

To see this,
fix $s\in \mathbb N$, with $s\ge 1$. For $h=(h_0, h_1, ..., h_s,...)\in \H$, consider the parameter sequence,
\[ \phi_{NT,s}(h) := f_s^{0} + N^{-1/2} \tf_s = f_s^{0}+ N^{-1/2} \sigma_s h_s, \quad \phi_{NT,s}(0) =f_s^{0}  \]
so $\sqrt{N} [\phi_{NT,s}(h)-\phi_{NT,s}(0)] = \sigma_s h_s$.
If we define $\dot \phi_s (h)= \sigma_s h_s$, then
\[ \sqrt{N} [\phi_{NT,s}(h)-\phi_{NT,s}(0)] = \dot \phi_s(h). \]
 Since $\dot \phi_s$ is  a coordinate projection map (multiplied by a positive constant $\sigma_s$), it is a continuous linear map, $\dot \phi_s: \H \rightarrow \R^{r}$. Its adjoint map $\dot \phi_s^*: \R^{r} \rightarrow \H$ (both spaces are self-dual) is the inclusion map (i.e., embedding): $\dot \phi_s^* x= (0,...,0, \sigma_s x, 0,...)\in \H$, for all $x\in \R^{r}$. The adjoint map satisfies
$\langle \dot \phi_s^* x, h\rangle =\sigma_s x' h_s = x' \dot \phi_s(h)=\langle x, \dot \phi_s(h)\rangle$.  Let $Z$ denote the limiting distribution of
efficient estimators of $f_s^{0}$.
Theorem 3.11.2 in van der Vaart and Wellner (\cite{vanderVaartWellner1996}, p.414) show that $x'Z\sim \N(0, \|\dot \phi_s^*x\|_{\H}^2)$ for all $x\in \R^{r}$. But
$\|\dot \phi_s^*x\|_{\H}^2 = \sigma_s^2 x'x$. It  follows that $Z \sim \N(0,\sigma_s^2 I_{r})$.
Thus the efficiency bound for regular estimators of $f_s^{0}$ is $\sigma_s^2 I_{r}$.

For $s=0$, the same argument shows that the efficiency bound for regular estimators of $\alpha^0$ is $1/(\gamma+\nu)$. In summary, we have

\begin{corollary} \label{coro:coro5} Under the assumptions of Theorem \ref{thm:localLR-2}, the asymptotic efficiency bound for regular estimators of $\alpha^0$
is $1/(\gamma+\nu)$, and the efficiency bound for regular estimators of $f_t^{0}$ is  $\sigma_t^2 I_{r}$.
\end{corollary}

It can be shown that the efficiency bound $1/(\gamma+\nu)$ corresponds to the case in which the incidental parameters $F^0=(f_1^0,f_2^0,...,f_T^0)'$ are known, thus the implied efficiency bound is likely too low to be attainable.
The implication is that the $\ell^2$ perturbation is ``too small''. Intuitively, the smaller the local parameter space, the lower the efficiency bound, making it  harder to achieve the implied bound. Interestingly, however, the $\ell^2$
local parameter space is appropriately sized for the fixed effects estimator, where both loadings and factors are estimated, as shown by \cite{IwakuraOkui2014}.

However, if $ F^0$ satisfies Assumption C, as in Theorem \ref{thm:localLR-1new}, then  Lemma \ref{lemma:11} in the appendix shows that
 $ \nu = 0$. As a result, the efficiency bound simplifies to $1/\gamma$. This means that $\ell^2$ is a suitable local parameter space for the maximum likelihood estimator when combined with Assumption C.  In this case, the efficiency bound is attainable by  the QMLE.

\section{Conclusion}

We derive the efficiency bound under normality for estimating dynamic panel models with interactive effects
 by formulating the model as a simultaneous equations system.
We show that quasi-maximum likelihood method applied to the system  attains the efficiency bound. These results are obtained under an increasing number of incidental parameters.


\begin{appendix}
\section*{Proof of Results} \label{appn}

We first consider the local parameter space for $\tf \in \ell_r^\infty$.
Let $G=F+\rootNT \tF$. We drop the superscript 0 associated with true parameters to make the notation less cumbersome.
 When evaluated at the true parameters, the likelihood function is, see (\ref{eq:concen-qmle})
\begin{equation} \label{eq:concen-qmle1} \ell(\theta^0) =-\frac N 2 \log|FF'+D|-\frac 1 2 \sumiN (y_i-\bar y)'B'(FF'+D)^{-1} B(y_i-\bar y) \end{equation}
When evaluated at the local parameters

\begin{align*} \ell(\theta^0+\rootNT \tilde \theta) & = -\frac N 2 \log|GG'+D| \\
& -\frac 1 2 \sumiN (y_i-\bar y)'(B -\talpha \rootNT J)' (GG'+D)^{-1} (B -\talpha \rootNT J)  (y_i-\bar y) \\
& =  -\frac N 2 \log|GG'+D|-\frac 1 2 \sumiN (y_i-\bar y)'B'(GG'+D)^{-1} B (y_i-\bar y)\\
& + \ta \rootNT \sumiN (y_i-\bar y)'J'(GG'+D)^{-1} B(y_i-\bar y)\\
& -\frac 1 2 \ta^2 \frac 1 {NT} \sumiN (y_i-\bar y)'J'(GG'+D)^{-1} J (y_i-\bar y)
\end{align*}
Thus, the difference is
\begin{align}   \ell(\theta^0+\rootNT \tilde \theta)& -\ell(\theta^0)  =  \notag\\
- & \frac N 2 \Big[\log|GG'+D|-\log|FF'+D| \Big]   \notag \\
- & \frac 1 2\sumiN  (y_i-\bar y)'B' \Big[(GG'+D)^{-1}-(FF'+D)^{-1}\Big] B (y_i-\bar y) \label{eq:LR1}\\
 + &  \ta \rootNT \sumiN (y_i-\bar y)'J'(GG'+D)^{-1} B(y_i-\bar y) \notag \\
 - & \frac 1 2 \ta^2 \frac 1 {NT} \sumiN (y_i-\bar y)'J'(GG'+D)^{-1} J (y_i-\bar y)  \notag
 \end{align}
 Notice that $ B(y_i-\bar y)= F(\lambda_i-\bar \lambda) + \eps_i-\bar \eps$, where $\bar \lambda =\frac 1 N \sumiN \lambda_i$ and $\bar \eps=\frac 1 N \sumiN \eps_i$. Furthermore, from $y_i-\bar y= B^{-1}  F(\lambda_i-\bar \lambda) + B^{-1}( \eps_i-\bar \eps)$, right multiply by $J$ and notice $J B^{-1}= L$ (see equation (\ref{JL})), we have
 $J(y_i-\bar y) =L F(\lambda_i-\bar \lambda) +L( \eps_i-\bar \eps)$. We also write $J(y_i-\bar y) = y_{i,-1} -\bar y_{-1},$  where
 $y_{i,-1}=(0, y_{i1},...,y_{i,T-1})'$ and $ \bar y_{-1}=(0, \bar y_1,...,\bar y_{T-1})'$.
 Thus we can rewrite the log-likelihood ratio as
\begin{align}   \ell(\theta^0+\rootNT \tilde \theta)& -\ell(\theta^0)  =  \notag\\
- & \frac N 2 \Big[\log|GG'+D|-\log|FF'+D| \Big]   \notag \\
- & \frac 1 2\sumiN  \Big[ F(\lambda_i-\bar \lambda) + \eps_i-\bar \eps\Big]'  \Big[(GG'+D)^{-1}-(FF'+D)^{-1}\Big] \Big[ F(\lambda_i-\bar \lambda) + \eps_i-\bar \eps\Big]   \label{eq:LR1}\\
 + &  \ta \rootNT \sumiN (y_{i,-1}-\bar y_{-1})'(GG'+D)^{-1} \Big[ F(\lambda_i-\bar \lambda) + \eps_i-\bar \eps\Big]  \notag \\
 - & \frac 1 2 \ta^2 \frac 1 {NT} \sumiN (y_{i,-1}-\bar y_{-1})'(GG'+D)^{-1}  (y_{i,-1}-\bar y_{-1}).  \notag
 \end{align}
 We shall derive the limit of the log likelihood ratio.
 Throughout, we use the matrix inversion formula (Woodbury formula)
\[ (FF'+D)^{-1}=D^{-1} -D^{-1}F(I_r+F'D^{-1} F)^{-1} F'D^{-1} \]
\[ (FF'+D)^{-1}F = D^{-1}F(I_r + F'D^{-1}F)^{-1}  \]
and the matrix determinant result
\[  |FF'+D|=|D||I_r +F'D^{-1}F| \]
From now on, we assume $r=1$ to simplify  the derivation.

We  define $\omega_F^2$ and $\eta_F^2$ as
\[ \omega_F^2=\frac 1 T  F'D^{-1} F, \quad \eta_F^2 =\frac 1 T (1+F'D^{-1} F) = \frac 1 T+ w_F^2   \]

\begin{lemma} \label{lemma:1} For $G= F +\rootNT \tF$,
\begin{equation}\label{log-remainder}
 -\frac N 2 \Big[\log|GG'+D|-\log|FF'+D| \Big]
 = -\sqrt{\frac N T} (F'D^{-1}\tF/T) \frac 1 {\eta_F^2} + O(\frac 1 T)  \end{equation}
 \end{lemma}
Proof:  Notice
\begin{equation*} \begin{split}  |GG'+D| = &   |(\FplustF)(\FplustF)'+D|  \\
 = & |D| [1+ (\FplustF)'D^{-1}(\FplustF)] \\
 = & |D|\Big( 1+F'D^{-1}F + 2 \rootNT F'D^{-1}\tF + \frac 1 {NT} \tF'D^{-1}\tF \Big) \\
 = & |D| (1+F'D^{-1}F) \Big[ 1+ 2 \frac 1 T \rootNT F'D^{-1}\tF \frac 1 {\eta_F^2} + \frac 1 {NT^2} \tF'D^{-1}\tF \frac 1 {\eta_F^2} \Big]   \\
 \end{split} \end{equation*}
  From $\log(1+x)= x +O(x^2)$ for small $x$ (big $O$)
\[ \log|GG'+D|-\log|FF'+D| = \log\Big[ 1+ 2 \frac 1 T \rootNT F'D^{-1}\tF/\eta_F^2 + \frac 1 {NT^2} \tF'D^{-1}\tF/\eta_F^2 \Big] \]
\[ =2 \frac 1 T \rootNT F'D^{-1}\tF/\eta_F^2 + \frac 1 {NT^2} \tF'D^{-1}\tF/\eta_F^2 +O(\frac 1 {NT})\]
where $O(x^2)=O(1/(NT))$.
Thus
\begin{align*} & -\frac N 2 \Big[\log|GG'+D|-\log|FF'+D| \Big] \\
 & = - \frac N T \rootNT F'D^{-1}\tF/\eta_F^2 - \frac 1 {2T^2} \tF'D^{-1}\tF/\eta_F^2 + O(\frac 1 T)
 \end{align*}
The second term on the right is also $O(1/T)$. This proves Lemma \ref{lemma:1}. $\Box$

\begin{lemma} \label{lemma:2} Let $H:=  (1+G'D^{-1}G)^{-1} -(1+F'D^{-1}F)^{-1}$. Then
\begin{align}
\label{H} H = &- 2 \frac 1 {T^2} \rootNT F'D^{-1}\tF \frac {1}{\eta_F^4}-\frac 1 {NT^3} \tF'D^{-1}\tF \frac {1}{\eta_F^4}  \\
  &  +4(\frac{F'D^{-1}\tF} T)^2 \frac {1} {NT^2\eta_F^6} + O(\frac 1 {N^{3/2}T^{5/2}})  \nonumber \end{align}
\end{lemma}
Proof:
\begin{align*}
 (1+G'D^{-1}G)& =
\Big(1+F'D^{-1}F + 2 \rootNT F'D^{-1}\tF + \frac 1 {NT} \tF'D^{-1}\tF \Big) \\
 & =(1+F'D^{-1}F) \Big( 1+ \frac{ 2  \rootNT F'D^{-1}\tF + \frac 1 {NT} \tF'D^{-1}\tF }{1+F'D^{-1}F}  \Big)  \\
 (1+G'D^{-1}G)^{-1}
 & =(1+F'D^{-1}F)^{-1} \Big( 1+ \frac{ 2  \rootNT F'D^{-1}\tF + \frac 1 {NT} \tF'D^{-1}\tF }{1+F'D^{-1}F}  \Big)^{-1}
 \end{align*}
Let $A= 2  \rootNT F'D^{-1}\tF + \frac 1 {NT} \tF'D^{-1}\tF$ and use the expansion $1/(1+x) =1-x +x^2 +O(x^3)$ for small $x$, we have
\begin{align*} & (1+G'D^{-1}G)^{-1}\\
& = (1+F'D^{-1}F)^{-1} \Big( 1 -\frac A {1+F'D^{-1}F} +\frac {A^2}{(1+F'D^{-1}F)^2}   + O(\frac { A^3} {(1+F'D^{-1}F)^3}) \Big) \\
& =(1+F'D^{-1}F)^{-1}  -\frac A {(1+F'D^{-1}F)^2} +\frac {A^2}{(1+F'D^{-1}F)^3 } + O(\frac { A^3} {(1+F'D^{-1}F)^4}) \\
\end{align*}
Now
\begin{align*}
  -\frac A {(1+F'D^{-1}F)^2} & = - 2 \frac 1 {T^2} \rootNT F'D^{-1}\tF \frac {1}{\eta_F^4}-\frac 1 {NT^3} \tF'D^{-1}\tF \frac {1}{\eta_F^4} \\
 \frac {A^2}{(1+F'D^{-1}F)^3 } & =\frac {A^2}{T^3 \eta_F^6} = 4(\frac{F'D^{-1}\tF} T)^2 \frac {1} {NT^2\eta_F^6} +O(\frac 1 {N^{3/2} T^{5/2}}) \\
 \frac {A^3}{(1+F'D^{-1}F)^4 } & = O(\frac 1 {N^{3/2}T^{5/2}}) \end{align*}
This proves the lemma. $\Box$.

A remark is in order.  In the above analysis, if we write $A=a+b$, then $A^2 =a^2 +2 ab +b^2$. Only $a^2/(T^3\eta_F^6)$ is non-negligible, $ab$ and $b^2$ are of smaller order of magnitude.
Finally, $A^3/(T^4\eta_F^8)$ is also a smaller magnitude.

\begin{lemma} \label{lemma:3} The following $T\times T$ matrix $\Xi$  satisfies
\[  \Xi:=(GG'+D)^{-1}- (FF'+D)^{-1} = -\Xi_a-\Xi_b-\Xi_c-\Xi_d +R\]
where
\begin{align}
&\Xi_a=   H D^{-1} FF'D^{-1} \label{Xib} \\
& \Xi_b =\Big[ \rootNT
\Big( \frac 1 {T \eta_F^2}
 \Big)-2 \frac 1 {NT^3}F'D^{-1} \tF \frac 1 {\eta_F^4}  \Big] D^{-1} F \tF'D^{-1} \label{Xic}\\
& \Xi_c =\Big[ \rootNT
\Big( \frac 1 {T\eta_F^2}
 \Big)-2 \frac 1 {NT^3}F'D^{-1} \tF \frac 1{\eta_F^4}  \Big] D^{-1} \tF F'D^{-1} \label{Xid} \\
 &\Xi_d= \frac 1 {NT^2}  \frac 1{\eta_F^2}D^{-1} \tF\tF'D^{-1}\label{Xie}  \end{align}
 where $H$ is defined in Lemma \ref{lemma:2}, and $R$ satisfies
 $\|R\|_2 =O(1/(NT)^{3/2})$.
\end{lemma}
Proof: From the Woodbury formula
\[ -\Xi=  D^{-1}G(1+  G'D^{-1}G)^{-1}  G'D^{-1}-D^{-1}F(1+F'D^{-1}F)^{-1}F'D^{-1}  \]
By Lemma \ref{lemma:2}, we can write
\begin{equation}\label{GGGG}  G(1+G'D^{-1}G)^{-1}G'=(1+F'D^{-1}F)^{-1} GG' + H GG'  \end{equation}
where $H$ is defined in Lemma \ref{lemma:2}.
The first term on the right hand side above is
\begin{align*} (1+F'D^{-1}F)^{-1}GG' &=(1+F'D^{-1}F)^{-1}
\Big(FF'+\rootNT F\tF' + \rootNT \tF F'  +\frac 1 {NT} \tF\tF'\Big).
  \end{align*}
 Thus, pre- and post-multiplying by $D^{-1}$
\begin{align}
  (1+ & F'D^{-1}F)^{-1}   D^{-1} GG D^{-1}- (1+F'D^{-1}F)^{-1} D^{-1} FF' D^{-1}   \nonumber \\
 &= (1+F'D^{-1}F)^{-1}  \Big[\rootNT D^{-1} F\tF' D^{-1}  + \rootNT D^{-1} \tF F' D^{-1}   +\frac 1 {NT} D^{-1} \tF\tF' D^{-1} \Big] \nonumber\\
  &=\frac 1 { T\eta_F^2}  \rootNT D^{-1} F\tF' D^{-1}  + \frac 1 { T\eta_F^2} \rootNT D^{-1} \tF F' D^{-1}   +\frac 1 {NT^2\eta_F^2 } D^{-1} \tF\tF' D^{-1}
 \label{GG-FF}
\end{align}

Next, consider $ H  GG'$ in (\ref{GGGG}), using $GG'=FF'+\rootNT F \tF'+\rootNT \tF F'+\frac 1 {NT} \tF\tF' $,
\begin{equation}\label{HGG}  H  GG' = H  FF'+ H  \rootNT F \tF' +  H  \rootNT \tF F' +  H \frac 1 {NT} \tF \tF'  \end{equation}
where $ H $ is given in (\ref{H}).
All of the four terms in $H$  are non-negligible for the matrix $HFF'$; only the first term in $H$ is non-negligible  for  $H \rootNT F\tF'$ and  $H \rootNT \tF F'$.
More specifically, notice
\[ H = - 2 \frac 1 {T^2} \rootNT F'D^{-1}\tF \frac {1}{\eta_F^4} + O(\frac 1 {NT^2}) \]
Left multiplying $H$ by $\rootNT F \tF'$,
\begin{align*}  H \rootNT F \tF' & = - 2 \frac 1 {NT^3}  F'D^{-1}\tF \frac {1}{\eta_F^4}  F \tF' + O(\frac 1 {NT^2}) \rootNT F \tF' \\
& = - 2 \frac 1 {NT^3}  F'D^{-1}\tF \frac {1}{\eta_F^4}  F \tF' + R_1
\end{align*}
where $\|R_1||_2= O(1/( NT)^{3/2})$ because the spectrum norm of $F\tF'$ is $O(T)$. Similarly,
\begin{align*}  H \rootNT \tF F' & = - 2 \frac 1 {NT^3}  F'D^{-1}\tF \frac {1}{\eta_F^4}   \tF F' + R_2
\end{align*}
where $\|R_2||_2= O(1/( NT)^{3/2})$.  Finally write
\[ R_3 := H \frac 1 {NT} \tF \tF' \]
then $\|R_3 \|_2 = O(1/( NT)^{3/2})$ because $H=O(N^{-1/2} T^{-3/2})$, and $\|\tF\tF'\|_2=O(T)$.
Pre- and post-multiply both sides of (\ref{HGG}) by $D^{-1}$ to obtain
\begin{align}  H D^{-1} GG'D^{-1} & =H D^{-1} F F' D^{-1} \notag \\
& - 2 \Big(\frac 1 {NT^3}  F'D^{-1}\tF \frac {1}{\eta_F^4}\Big) D^{-1}  F \tF' D^{-1} \label{HDGGD}\\
& - 2 \Big(\frac 1 {NT^3}  F'D^{-1}\tF \frac {1}{\eta_F^4}\Big) D^{-1}  \tF F' D^{-1} \notag\\
&  +D^{-1}(R_1+R_2 +R_3)D^{-1}  \notag
\end{align}
Let $R=D^{-1}(R_1+R_2 +R_3)D^{-1} $, then $\|R\|_2=\|R_j\|_2= O(1/( NT)^{3/2})$ for $j=1,2,3$ since  $\|D^{-1}\|_2= O_p(1)$.
The sum of
(\ref{GG-FF}) and (\ref{HDGGD}) is equal to $-\Xi$. This proves the lemma. $\Box$.

\begin{lemma} \label{lemma:4}
\begin{align*} -\frac 1 2  \sumiN  & [F(\lambda_i-\bar \lambda)]'\Big[(GG'+D)^{-1}-(FF'+D)^{-1}\Big] F(\lambda_i-\bar \lambda)  \\
& =\sqrt{\frac N T} (F'D^{-1}\tF/T) \frac 1{\eta_F^2} - \frac 1 2 \frac 1 T \tF' \FFD \tF  \\
& +O_p( N^{1/2} {T^{-3/2}}) + O_p(1/T) +O_p( N^{-1/2}).
 \end{align*}
\end{lemma}

Note the first term has an opposite sign with (\ref{log-remainder}).

\noindent
Proof:
By Lemma \ref{lemma:3},
\begin{equation}\begin{split}
 \sumiN [F(\lambda_i-\bar \lambda)]' & \Xi_a  F(\lambda_i-\bar \lambda)  =    H \sumiN   [F(\lambda_i-\bar \lambda)]' D^{-1}FF'D^{-1} F(\lambda_i-\bar \lambda)\\
 = & H (T^2 \omega_F^4)  \sumiN (\lambda_i-\bar \lambda)^2 \\
 = & \bigg[ -2\sqrt{NT}(F'D^{-1}\tF/T) \frac {\omega_F^4}{\eta_F^4}\\
&  -(\tF'D^{-1}\tF/T) \frac {\omega_F^4}{\eta_F^4} \\
& +  4(F'D^{-1}\tF/T)^2  \frac {\omega_F^4}{\eta_F^6}\bigg]   \frac 1 N \sumiN (\lambda_i-\bar \lambda)^2 +O_p(\rootNT),
 \end{split} \end{equation}
and
\begin{align*} \sumiN  & [F(\lambda_i-\bar \lambda)]'(\Xi_b +\Xi_c)F(\lambda_i-\bar \lambda) \\
   &=  \left[ 2\sqrt{NT}(\tF'D^{-1}F/T)  \frac   {\omega_F^2}{\eta_F^2}
 - 4 (F'D^{-1}\tF/T)^2 \frac  {\omega_F^2}{\eta_F^4} \right ] \frac 1 N \sumiN (\lambda_i-\bar \lambda)^2. \end{align*}
 Adding the two expressions
 \begin{align*} \sumiN & [F(\lambda_i-\bar \lambda)]'  (\Xi_a+\Xi_b+\Xi_c)  F(\lambda_i-\bar \lambda)\\
 & = \left[  2\sqrt{NT}(F'D^{-1}\tF/T) [  \frac   {\omega_F^2}{\eta_F^2}- \frac {\omega_F^4}{\eta_F^4} ]
 -(\tF'D^{-1}\tF/T)\frac {\omega_F^4}{\eta_F^4}\right] \frac 1 N \sumiN (\lambda_i-\bar \lambda)^2\\
 & + \left[4(F'D^{-1}\tF/T)^2  \Big(\frac {\omega_F^4}{\eta_F^6}- \frac {\omega_F^2}{\eta_F^4}\Big) \right] \frac 1 N \sumiN (\lambda_i-\bar \lambda)^2.
  \end{align*}
Note
\[ \frac   {\omega_F^2}{\eta_F^2}- \frac {\omega_F^4}{\eta_F^4}\equiv \frac   {\omega_F^2}{\eta_F^4}\frac 1 T = \frac 1 {T \eta_F^2} + O(\frac 1 {T^2}). \]
Hence
\[  2\sqrt{NT}(F'D^{-1}\tF/T) [  \frac   {\omega_F^2}{\eta_F^2}- \frac {\omega_F^4}{\eta_F^4} ]  =   2\sqrt{N/T}(F'D^{-1}\tF/T)\frac 1 {\eta_F^2}  + O_p(N^{1/2} T^{-3/2}). \]
Using  the fact that  $\omega_F^j/\eta_F^k= 1+O(1/T)$ for $j,k=2,4,6$,
\[ (\tF'D^{-1}\tF/T)\frac {\omega_F^4}{\eta_F^4}= (\tF'D^{-1}\tF/T) + O(1/T), \]
\[ 4(F'D^{-1}\tF/T)^2  \Big(\frac {\omega_F^4}{\eta_F^6}- \frac {\omega_F^2}{\eta_F^4}\Big) = O(1/T). \]
Combined with $\frac 1 N \sumiN (\lambda_i-\bar \lambda)^2= 1+O_p(N^{-1/2})$, we obtain
\begin{align} \sumiN & [F(\lambda_i-\bar \lambda)]'  (\Xi_a+\Xi_b+\Xi_c)  F(\lambda_i-\bar \lambda) \nonumber \\
& =2\sqrt{N/T}(F'D^{-1}\tF/T)\frac 1 {\eta_F^2} -(\tF'D^{-1}\tF/T) \label{Flambda(a+b+c)Flambda}  \\
& + O_p(N^{1/2} T^{-3/2})  +O_p(\rootNT) +O(\frac 1 T).  \nonumber
\end{align}
Next,
\begin{align}  \sumiN  & [F(\lambda_i-\bar \lambda)]' \Xi_d F(\lambda_i-\bar \lambda)  \nonumber \\
& =(F'D^{-1}\tF/T)^2  \frac 1 {\eta_F^2} \frac 1 N \sumiN (\lambda_i-\bar \lambda)^2
= (F'D^{-1}\tF/T)^2  \frac 1 {\eta_F^2} + O_p(N^{-1/2}).  \label{Flambda(d)Flambda}
\end{align}
By summing  (\ref{Flambda(a+b+c)Flambda}) and (\ref{Flambda(d)Flambda}) and multiplying  by 1/2, we obtain
\begin{align*} -\frac 1 2 \sumiN & [F(\lambda_i-\bar \lambda)]' \Big[(GG'+D)^{-1}-(FF'+D)^{-1}\Big] F(\lambda_i-\bar \lambda)  \\
 & =\sqrt{\frac N T} (F'D^{-1}\tF/T) \frac 1{\eta_F^2} -\frac 1 2 (\tF'D^{-1}\tF/T) + \frac 1 2 (F'D^{-1}\tF/T)^2  \frac 1 { \eta_F^2} \\
 & + O_p(N^{1/2} T^{-3/2}) +O(1/T)+O_p(N^{-1/2})
 \end{align*}
The lemma is obtained by noting
 \begin{align*}  (\tF'D^{-1}\tF/T) &  -(F'D^{-1}\tF/T)^2  \frac 1 { \eta_F^2}=\frac 1 T \tF'(FF'+D)^{-1} \tF.
\end{align*}
 $\Box$.

\begin{lemma} \label{lemma:5}
\begin{equation} \label{eps-eps}  -\frac 1 2 \sumiN (\eps_i-\bar \eps)' \Big[(GG'+D)^{-1}-(FF'+D)^{-1}\Big] (\eps_i-\bar \eps) =\op \end{equation}
\end{lemma}
Proof: First note that $\bar \eps =\frac 1 N \sum_{k=1}^N \eps_k$. It can be shown that $\bar \eps$ is a dominated term and can be ignored. By the notation of Lemma \ref{lemma:3}, we evaluate $\sumiN \eps_i'\Xi \eps_i$. Note that $\Xi$ consists of four parts. For this lemma, it is sufficient to approximate $\Xi$ by $\Xi_1+\Xi_2+\Xi_3$, where
\begin{align}  \Xi_1 &=  \Big(2 \frac 1 {T^2} \rootNT F'D^{-1}\tF  \frac 1 {\eta_F^4}\Big)
D^{-1}FF'D^{-1} \nonumber \\ \label{Xi-appro}
\Xi_2 & =  -  \Big( \rootNT \frac 1 T  \frac 1 {\eta_F^2}\Big) D^{-1}F\tF'D^{-1}\\
\Xi_3 & =  -
\Big( \rootNT \frac 1 T  \frac 1 {\eta_F^2} \Big)D^{-1} \tF F'D^{-1} \nonumber
\end{align}

In the above approximation, we kept the first term of $H$ in (\ref{Xib}) ($H$ has four terms), and kept the very first term inside the brackets  in (\ref{Xic}) and (\ref{Xid}). All other terms are negligible in the evaluation of
$\sumiN \eps_i'\Xi\eps_i$.  Using trace, it is easy to obtain the expected values
\[ \E\sumiN \eps_i'\Xi_1\eps_i=  2 \sqrt{\frac N T} (F'D^{-1}\tF/T) \frac 1{\eta_F^2} \]
And
\[ \E\sumiN \eps_i'(\Xi_2+\Xi_3)\eps_i= - 2 \sqrt{\frac N T} (F'D^{-1}\tF/T) \frac 1{\eta_F^2} \]
Thus the sum of the expected values is zero. In addition, the deviation of each term from its expected value is  $o_p(1)$. This is because the variance of each term is
$O(1/T)=o(1)$. For example, $ \var(\sumiN \eps_i'\Xi_2\eps_i)=\tr(\Xi_2 D \Xi_2 D)+\tr(\Xi_2'D \Xi D)=\frac 1 T (\frac 1 T \tF'D^{-1}F)^2 \frac 1 {\eta_F^4}
+ \frac 1 T (\frac 1 T \tF'D^{-1}F)(\frac 1 T F'D^{-1}F) \frac 1 {\eta_F^4} =O(1/T) $.  We have used the fact that for normal $\eps_i\sim N(0,D)$,  $\var (\eps_i'A\eps_i)=\tr (A D A D)+\tr(A'D A D)$ for any $A$,
This proves the lemma. $\Box$.

\begin{lemma} \label{lemma:6}
\begin{align*}   \sumiN [F(\lambda_i-\bar \lambda)]'  & \Big[(GG'+D)^{-1}-(FF'+D)^{-1}\Big] (\eps_i-\bar \eps)\\
 & =   -  \rootNT \sumiN \lambda_i'\tF'\FFD \eps_i  +\op \end{align*}
\end{lemma}
{\bf  Proof:} Recall $\Xi=(GG'+D)^{-1}-(FF'+D)^{-1}$.
The preceding approximation of $\Xi$ by $\Xi_1+\Xi_2+\Xi_3$ in (\ref{Xi-appro}) is sufficient (other terms are negligible). We evaluate $\sumiN (F\lambda_i)'\Xi_k \eps_i$ $(k=1,2,3)$.  Here we ignore $\bar \lambda$ and $\bar \eps$, the associated terms are negligible.
\begin{align*}
\sumiN (F\lambda_i)'\Xi_1 \eps_i & = 2 (F'D^{-1}\tF/T)(F'D^{-1}F/T)\rootNT \sumiN F'D^{-1}\eps_i \lambda_i \frac 1 {\eta_F^4} \\
& =2 \frac  {\omega_F^2} {\eta_F^4} (F'D^{-1}\tF/T) \rootNT \sumiN F'D^{-1}\eps_i \lambda_i\\
& =2 \frac  1 {\eta_F^2} (F'D^{-1}\tF/T) \rootNT \sumiN F'D^{-1}\eps_i \lambda_i +O_p(1/T)
\end{align*}
\begin{align*}   \sumiN (F\lambda_i)'\Xi_2 \, \eps_i &
 = -\frac 1 {\eta_F^2}  (F'D^{-1}F/T)  \rootNT \sumiN \tF'D^{-1}\eps_i \lambda_i \\
& =- \rootNT \sumiN \tF'D^{-1}\eps_i \lambda_i+O_p(1/T)
\end{align*}
and
\[ \sumiN (F\lambda_i)'\Xi_3 \, \eps_i=  - \frac 1 {\eta_F^2} (F'D^{-1}\tF/T)  \rootNT \sumiN F'D^{-1}\eps_i \lambda_i\]

Combining the three expressions
\begin{align*}  \sumiN (F\lambda_i)'\Xi \eps_i & =-  \rootNT \sumiN \lambda_i' \tF'D^{-1}\eps_i +\frac 1 {\eta_F^2} (F'D^{-1}\tF/T)   \rootNT \sumiN  F'D^{-1}\eps_i\lambda_i +O_p(1/T) \\
& = -\rootNT \sumiN \lambda_i'\tF'(FF'+D)^{-1} \eps_i +O_p(1/T).
\end{align*}
This proves the lemma. $\Box$
\begin{corollary} \label{coro:2} Under the Assumptions of Theorem 1,
\begin{align*}
 -\frac 1 2 \sumiN [F(\lambda_i-\bar \lambda) +\eps_i-\bar \eps] & \Big[(GG'+D)^{-1}-(FF'+D)^{-1}\Big] [F(\lambda_i-\bar \lambda) +\eps_i-\bar \eps]\\
  & = \sqrt{\frac N T} (F'D^{-1}\tF/T) \frac 1{\eta_F^2}\\
 & + \rootNT \sumiN \lambda_i'\tF'\FFD \eps_i \\
& -\frac 1 2 \frac 1 T \tF' \FFD \tF  +\op  \end{align*}
\end{corollary}
Proof: This follows by combining the results of Lemmas \ref{lemma:4}, \ref{lemma:5}, and \ref{lemma:6}. $\Box$

\begin{lemma} \label{lemma:7}
\begin{align}  &  \rootNT   \sumiN (y_{i,-1}-\bar y_{-1})'  (FF'+D)^{-1} [F(\lambda_i-\bar \lambda) +\eps_i-\bar \eps] \notag  \\
& = \rootNT \sumiN (L\eps_i)'D^{-1}\eps_i
 + \rootNT \sumiN\lambda_i' (LF)'[\FFD ] \eps_i +o_p(1) \label{yFF_eps}
 \end{align}
\end{lemma}
Proof: Using $ y_{i,-1}-\bar y_{-1} =LF(\lambda_i-\bar \lambda) +L (\eps_i-\eps)$,  we have
\begin{align*} LHS &=  \rootNT  \sumiN[ LF(\lambda_i-\bar \lambda) +L (\eps_i-\eps)]'(FF'+ D)^{-1} F(\lambda_i-\bar \lambda) \\
 &   +\rootNT  \sumiN[ LF(\lambda_i-\bar \lambda) +L (\eps_i-\eps)]' (FF'+D)^{-1}(\eps_i-\bar \eps) \\
 &  :=a+b \end{align*}
where $a$ is defined as the frist term, $b$ as the second term.
From the  formula,
\[ (FF'+D)^{-1} F= D^{-1} F ( {1+F'D^{-1}F})^{-1}=D^{-1}F (1+T\omega_F^2)^{-1} \]
term $a$ equals
\begin{align*} a &= \frac {\sqrt{N/T}}  {1+T\omega_F^2} [\frac 1 N \sumiN (\lambda_i-\bar \lambda)^2] (F'L'D^{-1} F)  + \frac 1 {1+T\omega_F^2} \rootNT \sumiN (\eps_i-\bar \eps)'L'D^{-1} F (\lambda_i-\bar \lambda) \\
& = \frac {\sqrt{N/T}}  {1+T\omega_F^2}  (F'L'D^{-1} F) +O_p(N^{-1/2})+O_p(1/T)
\end{align*}
where we used $\frac 1 N \sumiN (\lambda_i-\bar \lambda)^2=1+O_p(N^{-1/2})$, and the second  term is $O_p(1/T)$.

For term $b$, we shall ignore $\bar \lambda$ and $\bar \eps$ (the  associated terms are negligible). By the Woodbury formula,
\begin{align*}
 b & = \rootNT \sumiN (LF \lambda_i)' D^{-1} \eps_i + \rootNT \sumiN (L\eps_i)'D^{-1} \eps_i \\
& - \rootNT \sumiN \frac {\lambda_i' F'L'D^{-1}F}{(1+F'D^{-1}F)}   F'D^{-1}\eps_i-\frac 1 {(1+F'D^{-1}F)} \rootNT \sumiN \eps_i' L' D^{-1}F F'D^{-1}\eps_i
 \end{align*}
Note the expected value of the last term in the preceding equation is
\[ -\frac {F'L'D^{-1} F}{1+F'D^{-1}F} \sqrt{\frac N T} \]
and the deviation  from its expected value is negligible (because its variance is $O(1/T)=o(1)$, following from the same argument in Lemma \ref{lemma:5}). The above expected value cancels out with term $a$. Thus, we can rewrite
\begin{align*} a+ b & =  \rootNT \sumiN (L\eps_i)'D^{-1} \eps_i +\rootNT \sumiN\lambda_i' (LF)' (FF'+D)^{-1} \eps_i + o_p(1)
\end{align*}
proving  Lemma \ref{lemma:7}.

\begin{lemma}\label{lemma:8}

\begin{align}  \rootNT \sumiN  & (y_{i,-1}-\bar y_{-1})'   \Big[(GG'+D)^{-1}-(FF'+D)^{-1}\Big] [F(\lambda_i-\bar \lambda)+ \eps_i-\bar\eps]  \label{yXi_eps} \\
&=- \frac 1 T (LF)'[\FFD ] \tF +o_p(1) \notag  \end{align}
\end{lemma}
Proof: It is not difficult to show
\[ \rootNT \sumiN (y_{i,-1}-\bar y_{-1})' \Big[(GG'+D)^{-1}-(FF'+D)^{-1}\Big] ( \eps_i-\bar \eps) =\op\]
we thus focus on $\rootNT \sumiN (y_{i,-1}-\bar y_{-1})' \Xi F(\lambda_i-\bar \lambda)$, where $\Xi=(GG'+D)^{-1}-(FF'+D)^{-1}$.
Approximating $\Xi$ by (\ref{Xi-appro}) is sufficient.
Using $ y_{i,-1}-\bar y_{-1} =LF(\lambda_i-\bar \lambda) +L (\eps_i-\eps)$,
\begin{align*}  & \rootNT \sumiN (y_{i,-1}-\bar y_{-1})' \Xi_1 F(\lambda_i-\bar \lambda) \\
 & = \rootNT \sumiN   [LF(\lambda_i-\bar \lambda) +L (\eps_i-\eps)]' \Big(2\frac 1 {T^2}\rootNT (F'D^{-1}\tF) \frac 1 {\eta_F^4}\Big)D^{-1}(FF') D^{-1} F(\lambda_i-\bar \lambda)\\
& = 2 (F'L'D^{-1}F/T) (F'D^{-1}\tF/T) \frac 1 { \eta_F^2} +o_p(1) \end{align*}
where the term involving $L(\eps_i-\bar \eps)$ is $O_p(N^{-1/2})$, thus negligible.
 Here we have used $F'D^{-1}F=T\omega_F^2$,  $\omega_F^2/\eta_F^4=1/\eta_F^2 +O(T^{-1})$, and  $\frac 1 N \sumiN (\lambda_i-\bar \lambda)^2=1+o_p(1)$.
Next,
\begin{align*}
& \rootNT \sumiN (y_{i,-1}-\bar y_{-1})' \Xi_2 F(\lambda_i-\bar \lambda) \\
 &=- \rootNT \sumiN  [LF(\lambda_i-\bar \lambda) +L (\eps_i-\eps)]' \Big( \rootNT \frac 1 T  \frac 1 {\eta_F^2}\Big) D^{-1}(F\tF') D^{-1}F(\lambda_i-\bar \lambda)\\
  &=-\frac 1 {\eta_F^2} (F'L'D^{-1}F/T)(\tF'D^{-1}F/T) +\op, \end{align*}
  where term involving  $L(\eps_i-\bar \eps)$ is negligible. Next
\begin{align*}
& \rootNT \sumiN (y_{i,-1}-\bar y_{-1})' \Xi_3 F(\lambda_i-\bar \lambda)\\
& =-\rootNT \sumiN [LF(\lambda_i-\bar \lambda) +L (\eps_i-\eps)]' \Big( \rootNT \frac 1 T  \frac 1 {\eta_F^2}\Big)D^{-1} (\tF F')D^{-1} F(\lambda_i-\bar\lambda) \\
& =-   (F'L'D^{-1}\tF/T) +\op \end{align*}
which follows from the same reasoning for the term involving $\Xi_1$.
Summing up,
\begin{align*}  & \rootNT \sumiN (y_{i,-1}-\bar y_{-1})' \Xi [F(\lambda_i-\bar \lambda)+ \eps_i-\bar \eps] \\
&=\frac 1 {\eta_F^2} (F'L'D^{-1}F/T)(\tF'D^{-1}F/T) -(F'L'D^{-1}\tF/T) +\op \\
& =-\frac 1 T (LF)'(FF'+D)^{-1} \tF+o_p(1)
 \end{align*}
proving the lemma.

\begin{corollary} \label{coro:3}
\begin{equation}  \begin{split}
 \talpha \rootNT \sumiN & (y_{i,-1}-\bar y_{-1})'   (GG'+D)^{-1}[F(\lambda_i-\bar \lambda)+ \eps_i-\bar \eps]  \\
& = \talpha \, \rootNT \sumiN
(L\eps_i)'D^{-1}\eps_i  \\
& +\talpha \, \rootNT \sumiN\lambda_i' (LF)'[\FFD ] \eps_i
\label{yGGFlambda_eps} \\
  &  -\talpha \,  \frac 1 T (LF)'[\FFD ] \tF +o_p(1)
  \end{split}
 \end{equation}
\end{corollary}
Proof: adding and subtracting terms
\begin{align*}
 \talpha \rootNT \sumiN & (y_{i,-1}-\bar y_{-1})'   (GG'+D)^{-1}[F(\lambda_i-\bar \lambda)+ \eps_i-\bar \eps]  \\
&=\talpha \rootNT \sumiN  (y_{i,-1}-\bar y_{-1})'   (FF'+D)^{-1}[F(\lambda_i-\bar \lambda)+ \eps_i-\bar \eps]  \\
&+\talpha \rootNT \sumiN  (y_{i,-1}-\bar y_{-1})'   \Xi [F(\lambda_i-\bar \lambda)+ \eps_i-\bar \eps]
  \end{align*}
where, by  definition, $\Xi=(GG'+D)^{-1} - (FF'+D)^{-1}$.  The corollary follows from Lemmas \ref{lemma:7} and \ref{lemma:8}, that is, by
summing (\ref{yFF_eps}) and (\ref{yXi_eps}),
where every term is multiplied by $\tilde \alpha$.  $\Box$

\begin{lemma} \label{lemma:9}
\begin{align}  \frac 1 {NT}    \sumiN & (y_{i,-1}-\bar y_{-1})'  (GG'+D)^{-1} (y_{i,-1}-\bar y_{-1}) \label{yGGy-final} \\
& = \frac 1 T \tr(L'D^{-1}LD )
 + \frac 1 T \tr[ (LF)' [\FFD  ](LF) ] +o_p(1)  \nonumber
  \end{align}
  \end{lemma}
  Proof:   Here it is sufficient to approximate $(GG'+D)^{-1}$ by $ (FF'+D)^{-1}$ (recall $(GG'+D)^{-1}=(FF'+D)^{-1}+\Xi$, terms involving $\Xi$ are negligible because of the factor $(NT)^{-1}$).
Using $ y_{i,-1}-\bar y_{-1} =LF(\lambda_i-\bar \lambda) +L (\eps_i-\bar \eps)$,   we  rewrite
\begin{align*} \frac 1 {NT}& \sumiN (y_{i,-1}-\bar y_{-1})'  (FF'+D)^{-1} (y_{i,-1}-\bar y_{-1})\\
& =\frac 1 {NT} \sumiN[LF(\lambda_i-\bar \lambda) +L (\eps_i-\bar \eps)]'(FF'+D)^{-1}[LF(\lambda_i-\bar \lambda) +L (\eps_i-\bar \eps)] \\
& = \frac 1 {NT} \sumiN[LF(\lambda_i-\bar \lambda)]'(F'F+D)^{-1} LF(\lambda_i-\bar \lambda)\\
& + 2 \frac 1 {NT} \sumiN [LF(\lambda_i-\bar \lambda)]'(F'F+D)^{-1} L (\eps_i-\bar \eps)\\
& + \frac 1 {NT} \sumiN [L (\eps_i-\bar \eps)]'(FF'+D)^{-1} L (\eps_i-\bar \eps) = a +b +c
\end{align*}
where the three terms are denoted by $a,b,c$, respectively. First,
\begin{align*}  a & = \frac 1 T \tr[ (LF)'(FF'+D)^{-1} LF \frac 1 N \sumiN(\lambda_i-\bar \lambda) (\lambda_i-\bar \lambda)' ]\\
& =\frac 1 T \tr[ (LF)'(FF'+D)^{-1} LF]
+O_p(N^{-1/2}) \end{align*}
where the second equality is due to $N^{-1}\sumiN(\lambda_i-\bar \lambda) (\lambda_i-\bar \lambda)'=I_r+O_p(N^{-1/2})$.

Term $b$ can be shown to be $O_p(1/\sqrt{NT})$ and thus negligible. For $c$, we use the Woodbury formula,
\begin{align*}  c & = \frac 1 {NT} \sumiN (\eps_i-\bar \eps)'L' D^{-1}L (\eps_i-\bar \eps)\\
& -\frac 1 {(1+F'D^{-1}F)} \frac 1 {NT} \sumiN (\eps_i-\bar \eps)'L'D^{-1}F F'D^{-1}L (\eps_i-\bar \eps) =c1-c2.
\end{align*}
The mean of $c1$, $\E (c1)= [(N-1)/N]\frac 1 T\tr(L'D^{-1}LD)= \frac 1 T\tr(L'D^{-1}LD)+O(1/N)$. The deviation from the mean is negligible, because the variance
of $c1$ is $O(1/N)$. Thus
\[ c1= \frac 1 T\tr(L'D^{-1}LD) +o_p(1) \]
Consider $c2$. Its expected value
\[ \E (c2) =  \frac 1 {(1+F'D^{-1}F)} (\frac {N-1}N)  \frac 1 T \tr [ F'D^{-1} L DL' D^{-1} F] =O(1/T) \]
Because $c2$  is  nonnegative and its expected value converges to zero, we have $c2=o_p(1)$.  Thus the LHS of (\ref{yGGy-final}) is determined by
$a$ and $c1$. This proves the lemma. $\Box$

 {\bf Proof of Theorem \ref{thm:localLR-1}.} The local likelihood ratio is given by (\ref{eq:LR1}). Using
 Lemma \ref{lemma:1}, Corollary \ref{coro:2}, Corollary \ref{coro:3},  Lemma \ref{lemma:9} (multiplied by $-\frac 1 2 \ta^2$), we obtain
\begin{equation}  \label{LR-final-2}
 \ell(\theta^0+\rootNT \tilde \theta)  -\ell(\theta^0) = A(\ttheta)+ o_p(1)  \end{equation}
where
\addtocounter{equation}{1}
\begin{align*}
A(\ttheta) & =  \rootNT \sumiN \lambda_i'\tF'\FFD \eps_i\\
  & + \talpha \, \rootNT \sumiN (L\eps_i)'D^{-1}\eps_i  \\
  & +\talpha \, \rootNT \sumiN\lambda_i' (LF)'[\FFD ] \eps_i  \\
  &-\frac 1 2 \frac 1 T \tr[\tF'\FFD \tF ] \tag{\theequation} \label{eq:Atheta} \\
  & -\frac 1 2 \talpha^2 \tr \Big[ \frac 1 T (L'D^{-1}LD ) \Big]   \\
  & -\frac 1 2 \talpha^2  \tr \Big[ \frac 1 T  (LF)' \FFD  (LF)\Big]\\
    &  -\talpha \,  \frac 1 T \tr[ (LF)'\FFD  \tF].
   \end{align*}
Next, we show that $\FFD$ can be replaced by $\DMDFD$. In particular,
\begin{equation} \label{eq:AthetaDelta} A(\ttheta) =\Delta_{NT}(\tilde \theta)-\frac 1 2 \E[\Delta_{NT}(\tilde \theta)]^2 + O_p(T^{-1}) \end{equation}
where $\Delta_{NT}(\tilde \theta)$ is defined in Theorem 1.
The replacement ensures that the last four terms of  $A(\ttheta)$ are exactly the variance of the first three terms of $A(\ttheta)$  (multiplied by -1/2).
Let $\Upsilon= \FFD-\DMDFD$. We  first show, concerning the first term of $A(\ttheta)$,
\[ \rootNT \sumiN \lambda_i'\tF'\Upsilon \eps_i = O_p(1/T) \]
For simplicity, assume $r=1$, we can write $\Upsilon =  \frac  1 {T^2 \omega_F^2 \eta_F^2} D^{-1} F  F D^{-1}$.  Hence,
\[ \rootNT \sumiN \lambda_i'\tF'\Upsilon \eps_i =\frac  1 {T  \omega_F^2 \eta_F^2} (\frac {\tF'D^{-1}F} T) \rootNT \sumiN F' D^{-1} \eps_i\lambda_i =O_p(1/T). \]
The same analysis is applicable to the third term of $A(\ttheta)$, that is, $\rootNT \sumiN \lambda_i' (LF)'\Upsilon \eps_i = O_p(1/T)$. Consider the fourth term,
\[ \frac 1 T \tF'\Upsilon \tF = \frac  1 {T  \omega_F^2 \eta_F^2} (\frac {\tF'D^{-1}F} T) ( \frac {F'D^{-1} \tF} T) =O(\frac 1 T). \]
Each of the last two terms of $A(\ttheta)$ being $O(1/T)$ follows from the same analysis. This establishes (\ref{eq:AthetaDelta}).
The proof of Theorem 1 is complete. $\Box$

To prove Theorem \ref{thm:localLR-1new}, we need the following result.

\begin{lemma} \label{lemma:10} Under Assumption C,  we have
\[ \tag{a} \frac 1 T F'L'D^{-1} F \rightarrow \frac 1 {(1-\alpha)} \int_0^1 \psi(s)\psi(s)' /\sigma(s)^2 ds \]
\[  \tag{b} \frac 1 T F'L'D^{-1} LF \rightarrow \frac 1 {(1-\alpha)^2} \int_0^1 \psi(s) \psi(s)'/\sigma(s)^2 ds \]
\end{lemma}
This lemma extends Lemma 1 of Bai (2013), where the proof was much simpler.

Proof of (a). For simplicity, assume a single factor, and recall $F=(f_1,f_2,...,f_T)'$, and $D=\diag(\sigma_1^2,...,\sigma_T^2)$.
\[ (LF)'D^{-1} F = \sum_{k=0}^{T-2} \alpha^k \sum_{t=k+2}^{T-k} f_{t-k-1} f_t/\sigma_t^2  \]
For any $\epsilon>0$, choose a finite $K$ such that $|\sum_{k=0}^K \alpha^k -1/(1-\alpha)|<\epsilon$ and $|\alpha|^{K}\le \epsilon$. Under Assumption C,
$f_t =\psi(t/T)$ and $\psi(s)$ is continuous on [0,1], thus uniformly continuous on [0,1]. For the previous $\epsilon>0$, there is an $\delta>0$ such that
for all $|x-y|<\delta$, we have $|\psi(x)-\psi(y)|<\epsilon$.
\[  \frac 1 T (LF)'D^{-1} F = \sum_{k=0}^{K} \alpha^k \frac 1 T  \sum_{t=k+2}^{T-k} f_{t-k-1} f_t/\sigma_t^2  + \sum_{k=K+1}^{T-2} \alpha^k  \frac 1 T \sum_{t=k+2}^{T-k} f_{t-k-1} f_t/\sigma_t^2  \]
Note the second term is small. Since $f_t$ and $1/\sigma_t^2$ are bounded by assumption, the second term is bounded by
$|\alpha|^K M /(1-|\alpha|) = \epsilon M /(1-|\alpha|) \le \epsilon M'$. Consider the first term.
\[ \frac 1 T  \sum_{t=k+2}^{T-k}  f_{t-k-1} f_t/\sigma_t^2=  \frac 1 T  \sum_{t=k+2}^{T-k}  (f_{t-k-1}-f_t) f_t/\sigma_t^2 + \frac 1 T  \sum_{t=k+2}^{T-k}  f_t f_t/\sigma_t^2 \]
Again by the boundedness of  $f_t$ and $1/\sigma_t^2$, we have
\[  |\frac 1 T  \sum_{t=k+2}^{T-k}  (f_{t-k-1}-f_t) f_t/\sigma_t^2| \le M \frac 1 T  \sum_{t=k+2}^{T-k}  |f_{t-k-1}-f_t| =
 M \frac 1 T  \sum_{t=k+2}^{T-k} |\psi(\frac{t-k-1} T)-\psi(\frac t T)|\]
For large enough $T$, for all $k \le K$, $(k+1)/T \le (K+1)/T < \delta$, thus $|\psi(\frac{t-k-1} T)-\psi(\frac t T)|<\epsilon$, for all $t$ and all $k\le K$.
Thus, for all large $T$,
\[ \left| \sum_{k=0}^{K} \alpha^k \frac 1 T  \sum_{t=k+2}^{T-k} f_{t-k-1} f_t/\sigma_t^2- \sum_{k=0}^{K} \alpha^k \frac 1 T  \sum_{t=k+2}^{T-k} f_t f_t/\sigma_t^2\right| \le M \epsilon /(1-|\alpha|) \]
Next,
\begin{align*} \Big| \sum_{k=0}^{K} & \alpha^k \frac 1 T  \sum_{t=k+2}^{T-k} f_t f_t/\sigma_t^2 - \frac 1 {1-\alpha} \int_0^1 \psi(s)^2/\sigma(s)^2 ds \Big|\\
&  \le \left|\sum_{k=0}^{K}  \alpha^k \Big(\frac 1 T  \sum_{t=k+2}^{T-k} f_t f_t/\sigma_t^2 - \frac 1 T  \sum_{t=1}^{T} f_t f_t/\sigma_t^2\Big)  \right| \\
&   +  \left|\Big( \sum_{k=0}^{K}  \alpha^k -\frac 1 {1-\alpha}\Big) \Big( \frac 1 T  \sum_{t=1}^{T} f_t f_t/\sigma_t^2\Big)  \right| \\
&   +  \left| \frac 1 {1-\alpha}  \Big( \frac 1 T  \sum_{t=1}^{T} f_t f_t/\sigma_t^2 -\int_0^1 \psi(s)^2/\sigma(s)^2 \Big)  \right|
\end{align*}
The first term on the right hand side is bounded by $M/(1-|\alpha|) (2K+1)/T$, which is less than $\epsilon$ for large $T$.  The second term is bounded by $\epsilon M$ by the choice of $K$, and third term is bounded by $\epsilon$ for all large $T$. In summary, there exists an $M<\infty$, independent of $T$ such that for every $\epsilon>0$, and for all sufficiently large $T$, we have
\[ \left \| \frac 1 T F'L'D^{-1} F - \frac 1 {(1-\alpha)} \int_0^1 \psi(s)\psi(s)' /\sigma(s)^2 ds \right\|\le M \epsilon \]
This proves part (a).

Proof of (b). Notice
\[
\frac 1 T F'L' D^{-1} LF = \frac 1 T \sum_{t=2}^{T} \frac 1 {\sigma_t^2} \left( \sum_{j=2}^{t} \alpha^{t - j} f_{j - 1} \right)^2 \]
\begin{equation} \label{LFDLF} =\sum_{j=0}^{T-2} \sum_{k=0}^{T-2} \alpha^{j+k}  \left( \frac 1 T \sum_{t=\max(j,k)+2}^{T} \frac{f_{t - j - 1} f_{t - k - 1}}{\sigma_t^2}\right)
\end{equation}
Although we could make the proof rigorous by using the argument in part a), this would be repetitive.  We therefore focus on the key insight.
For  any  fix $j$ and $k$,
\[ \frac 1 T \sum_{t=\max(j,k)+2}^{T} \frac{f_{t - j - 1} f_{t - k - 1}}{\sigma_t^2} \rightarrow \int_0^1 \psi(s)^2/\sigma(s)^2 ds \]
which does not depend on $j$ and $k$. For large enough J and $K$,
$ \sum_{j=0}^J \sum_{k=0}^K \alpha^{j+k} =(\sum_{j=0}^J \alpha^j) (\sum_{k=0}^K \alpha^k)$, which is close to $1/(1-\alpha)^2$. The tail part of (\ref{LFDLF})  is negligible. This gives (b). $\Box$

\begin{lemma} \label{lemma:11} Under Assumption C,  and for $\tf \in C_f$, and $\tF=(\tf_1,...,\tf_T)'$.  we have
\[ \tag{a} \frac 1 T (LF)' [\DMDFD] LF \rightarrow 0 \]
\[ \tag{b} \frac 1 T (LF)' [\DMDFD] \tF \rightarrow 0 \]
\end{lemma}
Proof  of (a). By definition,
\begin{align*}  \frac 1 T   (LF)'[\DMDFD] & LF  = \frac 1 T F'L' D^{-1} LF \\
& - \frac 1 T (F'L'D^{-1}F) \Big(\frac{F'D^{-1}F}  T\Big)^{-1} \frac 1 T (F'D^{-1} LF). \end{align*}
Part (a) then follows from Lemma \ref{lemma:10} and $(F'D^{-1}F/T)^{-1} \rightarrow [\int_0^1 \psi(s)\psi(s)'/\sigma(s)^2 ds]^{-1}$.

Proof of (b). The LHS of (b) is equal to
\[ \frac { (LF)' D^{-1} \tF} T   - \Big(\frac{(LF)'D^{-1}F} T \Big)\Big( \frac{F'D^{-1}F} T \Big)^{-1} \Big( \frac{ F'D^{-1}\tF} T\Big).  \]
For the first term,
using the same argument for Lemma \ref{lemma:10} part (a), we can show
\[ \frac 1 T (LF)'D^{-1}\tF \rightarrow \frac 1 {1-\alpha} \int_0^1 \frac 1 {\sigma(s)^2} \psi(s) \tpsi(s)' ds, \]
which is equal to zero on $C_f$.
The second term also converges to zero on $C_f$,
\begin{equation} \label{eq:FDtF} \frac{ F'D^{-1}\tF} T \rightarrow \int_0^1 \frac 1 {\sigma(s)^2} \psi(s) \tpsi(s)' ds =0. \end{equation}
This proves the lemma. $\Box$

\vspace{0.1in}
{\bf Proof of Theorem \ref{thm:localLR-1new}}.  Since the local parameter space $\H=\R \times C_f$  is a subspace of $\ell^\infty$ for Theorem \ref{thm:localLR-1}, the proof of Theorem \ref{thm:localLR-1} holds on $\H$. However, some of the  expressions in Theorem \ref{thm:localLR-1} are  simplified under $\H$.

We begin by showing the first term of  $\Delta_{NT}(\tilde \theta)$ (see (\ref{eq:DeltaNT})) can be simplified
\begin{align*} \rootNT \sumiN  &\lambda_i'\tF'\DMDFD \eps_i \\
& = \rootNT \sumiN \lambda_i'\tF'D^{-1}\eps_i -\rootNT \sumiN \lambda_i'\tF' D^{-1} F ( F'D^{-1}F)^{-1} F' D^{-1}\eps_i
\end{align*}
We show that the  second term is $o_p(1)$. It suffices to show its variance converges to zero since the mean is zero.  Its variance is
\[ \tr [\frac 1 T (\tF'D^{-1} F)  (F'D^{-1} F)^{-1} (F'D^{-1}\tF)] \]
But on $C_f$,  (\ref{eq:FDtF}) holds.
Hence the said variance converges to zero on $C_f$.

Next we show the third  term of  $\Delta_{NT}(\tilde \theta)$ is $o_p(1)$. That is,
\[ \rootNT \sumiN\lambda_i' (LF)'[\DMDFD ] \eps_i  =o_p(1). \]
We also show its variance converges to zero. Its variance is given by
\[ \tr \Big[ \frac 1 T  (LF)' \DMDFD (LF)\Big]   \]
which converges to zero by Lemma \ref{lemma:11} part (a).
Next, consider the first term of $\E [ \Delta_{NT}(\tilde \theta)]^2$ (see (\ref{eq:varDeltaNT})) . We show
\begin{equation} \label{eq:tF(FFD)tF} \frac 1 T \tr[\tF'\DMDFD \tF ] = \frac 1 T \tr( \tF' D^{-1} \tF) + o(1). \end{equation}
By definition,
\[ \frac 1 T \tF'\DMDFD \tF =\frac 1 T \tF'D^{-1} \tF - \frac 1 T \tF'D^{-1} F (F'D^{-1}F)^{-1} F'D^{-1} \tF \]
The second term in the above equation converges to zero on $C_f$  is argued earlier. Hence
(\ref{eq:tF(FFD)tF}) holds.  The third term of $\E [ \Delta_{NT}(\tilde \theta)]^2$ is $o(1)$ by Lemma \ref{lemma:11} part (a). The  last term of $\E [ \Delta_{NT}(\tilde \theta)]^2$   converges to zero by Lemma \ref{lemma:11} part (b).
In  summary,  Theorem  \ref{thm:localLR-1}  is simplified to
\begin{align*}
  \Delta_{NT}(\tilde \theta) -\frac 1 2 \E [ \Delta_{NT}(\tilde \theta)]^2 &  = \rootNT \sumiN \lambda_i'\tF' D^{-1}\eps_i
    + \talpha \, \rootNT \sumiN (L\eps_i)'D^{-1}\eps_i  \\ &  -\frac 1 2 \frac 1 T \tr \Big[\tF'D^{-1} \tF \Big]
  -\frac 1 2  \talpha^2  \Big[ \frac 1 T \tr(L'D^{-1}LD ) \Big] +o_p(1).
  \end{align*}
This proves Theorem \ref{thm:localLR-1new}. $\Box$

\newcommand{\fst}{\frac 1 {\sigma_t^2}}

\vspace{0.1in}
{\bf Proof of Theorem \ref{thm:localLR-2}}. With respect to the local parameters $\tF$,
the proof of Theorem \ref{thm:localLR-1} only uses $\|T^{-1/2}\tF\|=O(1)$. If $\tf \in \ell_r^2$, then $\|\tF\|=O(1)$. The entire  proof of Theorem \ref{thm:localLR-1} holds with $T^{-1/2}\tF$ replaced by $\tF$. In particular, equations  (\ref{eq:DeltaNT}) and (\ref{eq:varDeltaNT})  hold  with $T^{-1/2}\tF$ replaced by $\tF$ (that is, omitting $T^{-1/2}$) due to $\|\tF\|=O(1)$. Notice $\tF$ appears in three places in (\ref{eq:DeltaNT}) and (\ref{eq:varDeltaNT}). We analyze each of them.  The first term of (\ref{eq:DeltaNT})
after replacing $T^{-1/2}\tF$ with $\tF$ is written as
\begin{align*} \rootN \sumiN \lambda_i'\tF' & \DMDFD \eps_i  =\rootN \sumiN \lambda_i'\tF' D^{-1} \eps_i\\
& - \tr[\tF'D^{-1}F  (F'D^{-1}F)^{-1} \rootN \sumiN F'D^{-1}\eps_i\lambda_i']\end{align*}
But the second term on the righthand side is $o_p(1)$, because it can be written as (ignore the trace)
\[ \rootT (\tF'D^{-1}F)   (F'D^{-1}F/T)^{-1} \rootNT  \sumiN F'D^{-1}\eps_i\lambda_i' \]
Now  $\|F'D^{-1}F/T]^{-1}\|=O(1)$,
 $\rootNT  \sumiN F'D^{-1}\eps_i\lambda_i'=\rootNT \sumtT \sumiN \frac 1 {\sigma^2} f_t\lambda_i' \eps_{it} =O_p(1)$, but
\begin{equation} \label{eq:toeplitz}  \|\rootT (\tF'D^{-1}F)\|=\|\rootT \sumtT \fst \tf_t f_t' \|\le \frac 1 a M \rootT \sum_{t=1}^T \|\tf_t\| =o(1) \end{equation}
we have used $\sigma_t^2\ge a>0$, and $\|f_t\|\le M$. To see  $\rootT \sum_{t=1}^T \|\tf_t\| =o(1)$ for $\tf\in \ell_r^2$, notice
$\tf_s\rightarrow 0$ as $s\rightarrow \infty$, and by the Toeplitz lemma, $\rootT \sum_{t=1}^{\sqrt{T}}\|\tf_t\|
\rightarrow 0$. By the Cauchy-Schwarz, $\rootT \sum_{t=\sqrt{T}+1}^T \|\tf_t\|  \le (\sum_{t=\sqrt{T}+1}^T \|\tf_t\|^2)^{1/2}
\le (\sum_{t=\sqrt{T}}^\infty \|\tf_t\|^2 )^{1/2}\rightarrow 0.$ (Also see \cite{IwakuraOkui2014} for a similar result.)
Thus,
\[ \rootN \sumiN \lambda_i'\tF'\DMDFD \eps_i=\rootN \sumiN \lambda_i'\tF' D^{-1} \eps_i +o_p(1). \]
The first term of (\ref{eq:varDeltaNT}) after replacing $\frac 1 T \tF$ by $\tF$ becomes (ignore the -1/2 and the trace),
\begin{align*} \tF'\DMDFD \tF  & =\tF' D^{-1} \tF -\tF' D^{-1}F ( F'D^{-1}F)^{-1} F'D^{-1}\tF \\
 & = \tF' D^{-1} \tF +o(1) \end{align*}
owing to $(F'D^{-1}F/T)^{-1}=O(1)$ and $\tF' D^{-1}F/T^{1/2}=o(1)$ due to (\ref{eq:toeplitz}).
The last term of (\ref{eq:varDeltaNT})  being $o(1)$  with  $\tF$  in place of  $\rootT \tF$ follows from the same argument.

Collecting the simplified and the non-negligible terms, we obtain the expressions in Theorem \ref{thm:localLR-2}. $\Box$

\end{appendix}

\end{document}